\documentclass[12pt,a4paper]{article}
\usepackage[latin1]{inputenc}
\usepackage{amsmath}
\usepackage{amsfonts}
\usepackage{amssymb}
\usepackage{slashed}
\usepackage{graphicx, subfig}
\usepackage{cite}

\def\PFGstripminus-#1{#1}%
\def\PFGshift(#1,#2)#3{\raisebox{#2}[\height][\depth]{\hbox{%
  \ifdim#1<0pt\kern#1 #3\kern\PFGstripminus#1\else\kern#1 #3\kern-#1\fi}}}%

\def\to{\rightarrow}
\def\beq{\begin{equation}}
\def\eeq{\end{equation}}
\def\beeq{\begin{eqnarray}}
\def\eeeq{\end{eqnarray}}
\def\be{\begin{eqnarray}}
\def\ee{\end{eqnarray}}
\def\beal{\begin{align}}
\def\eeal{\end{align}}

\def\centeron#1#2{{\setbox0=\hbox{#1}\setbox1=\hbox{#2}\ifdim
\wd1>\wd0\kern.5\wd1\kern-.5\wd0\fi
\copy0\kern-.5\wd0\kern-.5\wd1\copy1\ifdim\wd0>\wd1
\kern.5\wd0\kern-.5\wd1\fi}}
\def\ltap{\;\centeron{\raise.35ex\hbox{$<$}}{\lower.65ex\hbox{$\sim$}}\;}
\def\gtap{\;\centeron{\raise.35ex\hbox{$>$}}{\lower.65ex\hbox{$\sim$}}\;}

\def\PFGstripminus-#1{#1}%
\def\PFGshift(#1,#2)#3{\raisebox{#2}[\height][\depth]{\hbox{%
  \ifdim#1<0pt\kern#1 #3\kern\PFGstripminus#1\else\kern#1 #3\kern-#1\fi}}}%

\newcommand{\GeV}{{\rm\ GeV}}

\newcommand{\TeV}{{\rm\ TeV}}

\def\ds{\displaystyle}
\def\bea{\begin{array}{c}}
\def\ea{\end{array}}
\def\be{\begin{equation}\bea\ds}
\def\ee{\ea\end{equation}}

\begin{document}

\begin{flushright}\mbox{YITP-SB-11-46}\\
\mbox{CERN-PH-TH/2012-031}\end{flushright}

\begin{center}

{\large {\bf Three-particle templates for boosted Higgs}}
\end{center}

\begin{center}
 {Leandro G. Almeida}$^{a}$, {Ozan Erdo\u{g}an}$^{b}$, {Jos\'{e} Juknevich}$^c$, {Seung J. Lee}$^d$, {Gilad Perez}$^{c,e}$, {George Sterman}$^b$ \\

\vskip 8pt

$^{a}${\it \small Institut de Physique Theorique, CEA-Saclay, F-91191, Gif-sur-Yvette cedex, France}\\
\vspace*{0.3cm}

$^{b}$ {\it \small C.N. Yang Institute for Theoretical Physics\\
Stony Brook University, Stony Brook, New York 11794-3840, USA}\\

\vspace*{0.3cm}

   $^c$ {\it \small Department of Particle Physics and Astrophysics \\
   Weizmann Institute of Science, Rehovot 76100, Israel}\\  

\vspace*{0.3cm}

   $^d$ {\it \small Department of Physics,
Korea Advanced Institute of Science and Technology \\ 335 Gwahak-ro, Yuseong-gu, Daejeon 305-701,
Korea}

$^e$ {\it \small CERN, Theory Division, CH1211 Geneva 23, Switzerland}

\vspace*{0.853cm}\end{center}

\begin{abstract}

We explore the ability of three-particle templates
to distinguish color neutral objects from QCD background.
This method is particularly useful to identify the standard model Higgs, as well as other massive neutral particles.
Simple cut-based analysis in the overlap distributions of the signal
and background is shown to provide a significant rejection power. By
combining with other discriminating variables, such as planar flow,
and several variables that depend on the partonic template, three-particle templates are used to characterize the influence of gluon emission and color flow in collider events.
The performance of the method is discussed for the case of a highly boosted Higgs in association with a leptonically-decaying $W$ boson.  
\end{abstract}

\section{Introduction}\label{Sec:intro}

Processes  with 
top quarks, standard model Higgs and $W/Z$ jets  play  a  key  role  in  high  energy  collisions. 
They  allow  for  tests  of  perturbative  QCD and are important  backgrounds  for  more  exotic 
phenomena. Over the past few years, scenarios have been proposed in which these heavy particles are produced 
at large transverse momentum~\cite{Butterworthsparticles,Butterworth:2002tt,Butterworth:2008iy,Butterworth:2009qa,KKG1,KKG2,Fitzpatrick:2007qr,Agashe:2007zd,Lillie:2007ve,Agashe:2007ki,Nath:2010zj}.  At high enough $p_T$, 
their decay products will appear as heavy, collimated jets
\cite{tscet,Banfi:2007gu}.   Even such exotic final states, however,
will coexist with a substantial tail of the mass distribution of light-parton QCD jets
\cite{Ellis:2007ib,Almeida:2008tp}.   Thus it will generally be necessary 
to study jet substructure systematically to distinguish such a
signal. 

Precise  predictions  for  the substructure of such boosted jets  are severely 
limited  by  the  complexity  of  the final state: 
in general it will be difficult to find an experimental observable perfectly correlated with a parton shower history, including color flow and hadronization.
Nevertheless, a number of methods to analyze high-$p_T$ jets have been proposed.   Generally, these methods depend on differences in the substructure of light-parton QCD jets compared to those from particle decays.
Diagnostics to detect this difference include infrared safe jet shapes \cite{Thaler:2008ju,Almeida:2008yp,arXiv:1011.2268,arXiv:1108.2701,arXiv:1101.2905}, direct analyses of 
jet substructure~\cite{Butterworth:2008iy,Kaplan:2008ie,Kribs:2009yh,Chekanov:2010vc,arXiv:1010.5253,arXiv:1011.4523,arXiv:1006.3213,arXiv:1010.0676,arXiv:1008.2202,arXiv:1012.2077,arXiv:1010.3698,  arXiv:1102.0557,arXiv:1007.2221,arXiv:1104.1646,arXiv:1011.1493},
grooming methods to improve jet mass resolution by reducing jet contamination \cite{Pruning,Krohn:2009th,Soper:2010xk}, and matrix element methods \cite{arXiv:1102.3480}. Jet substructure techniques have also
been studied in the context of specific particle searches,
where they have been shown to extend substantially the
reach of traditional search techniques in a wide variety of
scenarios \cite{arXiv:1108.1183, arXiv:1107.3563,arXiv:1102.1012,arXiv:1012.2866,arXiv:1111.4530,arXiv:1111.1719,arXiv:1106.3076,arXiv:1012.1316,Sung:2009iq,Gallicchio:2010sw}. In addition, experiments have started to use boosted hadronic objects as a probe of new physics in data \cite{arXiv:1101.0070,arXiv:1106.5952,arXiv:1101.3002}(for a summary of the experimental and theoretical progress in jet substructure, see, for instance, \cite{Abdesselam:2010pt,arXiv:1110.3684}).

In Ref.\ \cite{Almeida:2010pa} a template overlap method was developed for the
 quantitative comparison of the energy  flow of observed jets at high-$p_T$ with the flow from selected sets (the templates) of
partonic states.  The template overlap method can be summarized as follows, adopting the notation of Ref.\  \cite{Almeida:2010pa}.   We denote a set of particles or
calorimeter towers in a physical (or event-generated) jet by $|j\rangle$.   
The jet, of course, is identified by some algorithm.
Any such jet is to be compared with a large set of configurations  $|f\rangle$,
which represent sets of partonic momenta $p_1,\dots ,p_N$ that would be identified as a jet  by the same algorithm
we apply to the physical jets.
For a given $j$, we determine the template state for which some measure, ${\cal F}(j,f)\equiv \langle f | j \rangle$,
is maximized.   Although this measure is in principle completely free, we choose below
 a Gaussian in energy differences within angular regions surrounding each of the partons of the template.
 When the energy flow of the state perfectly tracks that of a template, the measure is maximized at unity.
Any region of partonic phase space $\{f\}$, can define a template, although in this analysis
we will be interested in templates that can describe boosted Higgs decays, in terms of the energies
and invariant masses of its partons.

 Although applied to boosted Higgs in this paper, the template overlap
 approach  is very flexible and can be applied to a wide variety of particle processes for which theoretical models have been established.  
 For each set of templates with a definite number of particles, $N$, the overlaps provide us with a tool to match unequivocally arbitrary final states $j$ to partonic partners $f[j]$.  Once a ``peak template" $f[j]$ is found, we can use it to characterize the energy flow of the state, which gives additional information on the likelihood that it is signal or background.    Similarly, it is possible to analyze the template functions found in this way to further discriminate events.   We will give examples below.
 
 As was shown in \cite{Almeida:2010pa}, the application of these ideas is particularly straightforward for top jets.  Much of
the QCD background is characterized by two sub-jets, 
with very different energy flow from the three-parton templates in general.
Indeed, for a lowest order partonic QCD jet consisting of the original parton
plus one soft gluon, there is no template state from top decay that 
matches the energy flow.   This gives a fundamental discrimination, to which
we can add additional information from event shapes.  
 
In practice the  procedure outlined in \cite{Almeida:2010pa} is not yet optimal for 
disentangling a Higgs, as opposed to top, signal from a QCD background, because at lowest
order (corresponding to templates with $N=2$) both boosted Higgs and
QCD jets consist of two particles. A successful identification
strategy should also make use of observables that do not map
exclusively onto the minimum number of partons. We go a further
step in this direction by employing templates with three partons, with
a cut to ensure that they are well-separated in phase space. To the extent that physical final states reflect partonic energy flow in both signal and background, the population of three-particle templates found from a given sample reflects the short-distance dynamics that produced the corresponding physical states.

In the following section we review the definition of template overlap
used in this paper.  Next, in Sec.~3, we  introduce an efficient
procedure to construct the templates and maximize the overlap. The
discriminating power of several observables, including planar flow,
and several new partonic template variables is discussed in Sec.~4. We
then study Higgs tagging performance of template overlap in Sec.~5.
Finally, we conclude in Sec.~6, leaving a few explicit calculations to the appendices.

\section{The Template Overlap Method}\label{Sec:generalizing_template_overlap}

The method developed in~\cite{Almeida:2010pa} makes use of template
overlaps that are functionals of energy flow of jet
events. The templates are sets of partonic momenta $f=\{p_1,\dots
,p_N\}$, with
\beeq
\sum_{i=1}^N p_i=P\, ,\, \quad P^2=M^2\, , 
\eeeq
which represent the decay products of a signal of mass
$M$.  Here we have adopted the expectation that a good, if not the
  best, rejection power is obtained when we use the signal
  distribution itself to construct our templates (see~\cite{maldacena}).  For example, the lowest order templates for
Higgs have $N=2$ and for the top decay, $N=3$, with
phase space in the latter case restricted by the $W$ mass: $t\rightarrow b+W \rightarrow b + (q\bar q)$. The number of
particles in the templates is not necessarily fixed, and templates
with more than the minimum number of particles are possible. We will
find, however, that combining templates in the full phase space for $N=3$ 
and $N=2$
already delivers encouraging results for the Higgs.

In order to compare the energy flow of any given (physical or
event-generated) jet to that of the signal on the unit sphere, denoted
by $\Omega$, following the notation of Ref.~\cite{Almeida:2010pa}, we represent the template energy flow as
$d E{(f)}/d \Omega$ while that 
of the jet is denoted by $d E{(j)}/d \Omega$.   
We may then represent a general overlap functional $Ov(j,f)$ between a
jet and a template as  
\beq
Ov(j,f) = \langle j|f\rangle = {\cal{F}}\left[ \frac{d E{(j)}}{d \Omega} , \frac{d E{(f)}}{d \Omega}  \right]\, .
\eeq
Here, the functional ${\cal F}$ is in principle completely free, and
by construction it is
maximized by the corresponding template for a given $j$\@. The measure  for the matching
between $j$ and the template $f$ can be taken as the weighted
difference of their energy flows integrated over some specific region
in the phase space. In our analysis, we will take the functional ${\cal
F}$ as a Gaussian in energy differences within angular regions
surrounding each parton in the template, i.e. around directions of the
$N$~template momenta $p_i$, 
\beeq
Ov_{N}(j,p_1,\dots ,p_N)  =  
{ \rm{max}}_{\tau^{(R)}_N}\,
\exp\left[\, -\ \sum_{a=1}^N \frac{1}{2\sigma^2_a}\left(  \int d^2 \hat n\,
  \frac{d E{(j)}}{d^2 \hat n} \theta_N ( \hat n, \hat n_a^{(f)})  - E_a^{(f)}\, \right)^2
   \right] ,
  \label{overlap2}
\eeeq
where the direction of parton $a$ in the template is $\hat n_a$
and its energy is $E_a^{(f)}$. These energies set the widths~$\sigma_a$ of the
Gaussians; while the functions $ \theta_N (\hat n, \hat n_a^{(f)})$
restrict the angular integrals to (nonintersecting) regions with a
definite width
surrounding each of the partons of the template.
The corresponding state will be referred as the ``peak template" $f[j]$ for state $j$.

A new element of our analysis is to use
templates with more than the minimum number of particles. This gives us the ability to resolve details of jet substructure, facilitating the capture of possible gluon radiation in the heavy particle decay, while still eliminating contamination from pile-up and the underlying event.   

A key step for defining template overlap is to choose appropriately the set of template states over which the maximization is performed. Ideally, one would determine $Ov$ by maximizing over all possible template states. In practice, however,  one needs to introduce a discretization in the space of template states. In order to make sure that the maximum overlap found by this reduced set is very close to the true maximum, a large number of template states is needed, roughly of order $100^{3N-4}$.  In general, such a maximization procedure can be computationally intensive when the number of template particles, $N$, is large, but in section \ref{Sec:Analysys_overview} we present an efficient algorithm to perform this task.  We emphasize that, once generated, the same set of template states is used for all the data.

In summary,  the output of 
the peak template method for any physical state $j$ is 
the value of the overlap, $Ov(j,f)$, and also the 
identity of the template state $f[j]$ to which the best match is found.
As we shall see, this will be of particular
value when we apply our method to boosted Higgs.

\section{Analysis overview}\label{Sec:Analysys_overview}

Motivated by Ref.~\cite{Butterworth:2008iy}, we now describe the application of the   template method
 to the production of a Higgs boson in association with a $W$ boson, $p+p \to W+H$, followed by the dominant light Higgs boson decay, to two b-tagged jets,
including schemes for generating templates and for discretizing the data. We wish to investigate the tagging efficiency for this process, and the fake rates from the background process $p+p \to W+ jets$. We emphasize that the method we propose is quite general, and it can be used for other massive object searches as well.
 
 In this section, we describe the construction and use of templates with $N=3$ for Higgs decay.
Three particle templates will allow us to test the influence of gluon emission and color flow, through their effect on energy flow.   This will provide an additional tool to discriminate between QCD and Higgs jets.  For example, we expect soft radiation from the boosted color singlet Higgs to be concentrated between the $b$ and the $\bar b$ decay products. This is to be contrasted to a jet initiated by a light parton, whose color is correlated with particles in other parts of phase space, producing radiation in the gaps between those particles and the jet system. 
 In this way, the template for a given state can provide evidence on its origin.
 The template method has the advantage of not requiring any special algorithmic technique and allows us to consider more elaborate jet substructure observables, and we will look at a few specific kinds in the following sections.

\subsection{Matrix Element Optimization}
\label{meopt}

In the preliminary study of Ref.  \cite{Almeida:2010pa}, the template
states, $f=p_1,\dots ,p_N$, were generated evenly over all the phase
space by a brute force method. This approach was very
effective for the top at LO, $t\rightarrow
b+W\rightarrow b + (q\bar{q})$, where the pair invariant mass was
constrained to be $m_W$\@. However, the full, unconstrained 3-particle
phase space translates to an increase of computational runtime. If we have $N$-body templates described by $3N-4$ phase-space
variables and we divide each of these variables into, say, $\sim$100
bins, then we have $\sim 100^{3N-4}$ total bins in the absence of
additional kinematic constraints.  A coarser discretization might generate errors in the estimation of $Ov_N(j,f)$, when the
matrix element is changing rapidly. We here present a new scheme which
significantly reduces the number of the template states needed for the computation of $Ov_N(j,f)$.

A more efficient approach is to generate templates in phase space according to the size of squared matrix elements of signal events.    This will ensure that most signal events will match templates very closely.   In regions where matrix elements are small, they are changing slowly, so that rare signal events will be well matched even where the density of templates is lower.    Such matrix element weighting methods are simple to implement and allow one to deemphasize the uninteresting majority of phase space\footnote{In different contexts, such an approach has been applied to Tevatron data\cite{TeVatronMatrixElement}\@.}. The idea is to associate to each template $f$ a weight defined as the probability to produce and observe a template in a given model, labelled by the letter $\alpha$.  We can readily evaluate such a probability in the ideal situation where the resolution of the
detector is perfect. The probability is then given by the squared matrix
element, summed over final state colors and polarizations and averaged over initial ones,
\be
{\cal P} ({\bf p}|\alpha) = \frac{1}{\sigma_\alpha} d\Phi \,| M_\alpha({\bf p})|^2  \label{MCxsection}.
\ee
Here $M_\alpha$ is the matrix element, $d\Phi$ is the phase-space measure and $\sigma_\alpha$ is the total cross section. 

The generation of templates according to the cross section in Eq. (\ref{MCxsection}) involves three steps:
\begin{itemize}
\item A sampling Monte Carlo routine is used to sample the phase
  space, ${\{\bf x ^{(r)} \}}_{r=1}^{3 N-4}$ with $r$ running over
  independent kinematical variables, according to the probability distribution ${\cal P} ({\bf p}|\alpha) $.
\item The same algorithm computes the templates, i.e. the parton momenta $\{ p_i\}_{i=1}^{N}$, from the set of ${\{\bf x ^{(r)} \}}_{r=1}^{3 N-4}$. 
\item The matrix elements contain singularities in certain kinematic
  configurations. For example, those with final states of several jets
  are singular if the jets are 
  nearly collinear or if the energy of a
  jet approaches zero. Hence, after the momenta are
  generated, a call is made to a routine that is used to apply cuts to
  the generated templates. A phase space point that fails the cut is
  then rejected and the template is not evaluated, and the state
  reverts to a template with fewer particles. 
\end{itemize}

\subsection{Selection and Discretization of the Data} 
\label{event_discretization}

We generate events for $W^+ +H \to l^+ \nu_l\, b\bar b $ and $W^ + + jets \to l^+
\nu_l+jets$ in a configuration with large transverse momentum, using
{\sc Pythia 8.150}\cite{pythia8}, {\sc
  Sherpa 1.3.0}\cite{sherpa} (with CKKW matching \cite{ckkw}), and
{\sc Madgraph}\cite{madgraph} interfaced to {\sc Pythia 6}~\cite{pythia6} (with
MLM matching\cite{mlm}).    Jets are reconstructed using {\sc Fastjet}
\cite{fastjet}, and the anti-$k_T$ algorithm \cite{antikt} with large
effective cone size $R=0.4,0.7$\@.   In this paper, we have chosen
plausible values for $R$, based on a combination of physics input and
a trial-and-error, but have not attempted to optimize them
systematically.\footnote{An optimal jet radius would be a compromise
  between taking it large enough to include perturbative final-state
  radiation, and small enough to avoid too much contamination from the
  underlying event and initial-state radiation.}  For each event,   we
find the jet with the highest transverse momentum and impose a jet
mass window for the Higgs.   We choose the jet mass window to be  110
GeV $\le m_J \le$ 130 GeV, with our reference Higgs boson mass chosen
to be $m_H=120$ GeV,  and jet energy 950 GeV $\le P_0 \le$ 1050 GeV.  This gives us a set of final states $j$.

For any state $j$, we determine the measured (or Monte Carlo (MC) generated) energy distribution, $dE(j)/d\Omega$, in the physical $\theta$-$\phi$ plane with respect to the jet axis for each reconstructed jet, and
we discretize this data into a jet-energy configuration.
In our demonstration for the Higgs, we discretize the $\theta$-$\phi$ plane into cells of size $\Delta \theta=0.04$ and $\Delta \phi=0.1$ \footnote{In a typical experimental setup the energy is discretized according to the detector resolution, and each pair (row,col) corresponds to a specific cell in the calorimeter.
At the LHC experiments~\cite{cal}, for instance, electromagnetic calorimeter cell size (in $\eta$ and $\phi$) is of ${\cal O}\big(0.025\times0.025\big)$ and of ${\cal O}\big(0.1\times0.1\big)$ for hadronic calorimeter cells.}.
Next, we again assemble a table of energies $E(\rm row_m, column_n)$, where $\rm row_m$ and $\rm column_n$ are the row and column number corresponding to the discretized values of $\theta$ and $\phi$.

Before proceeding further, let us mention a few words concerning our choice of event selection criteria and the interpretation of our results.  Our focus in this work is to demonstrate that the combination of two- and three-particle templates can give a qualitative improvement in  the separation of Higgs signal events from QCD background. In order to simplify the analysis we have restricted our study to very large transverse momenta, not realistic at this time of the LHC running. However, we regard the findings described below as a proof of principle,   and we expect that more realistic parameters~\cite{WIP} ({for example, a lower  $p_T$ cut) and $b-$tagging will provide even higher discriminating power.

\subsection{Construction of template functions} \label{construction}

We define the leading order (next to leading order) templates in terms
of the lowest-order (next-to-lowest-order) decays of the Higgs, schematically,
\be
| f \rangle =
| h\rangle^{\rm (LO)} = | p_1,p_2\rangle
\, ,\ee
and
\be
| f \rangle =
| h\rangle^{\rm (NLO)} = | p_1,p_2,p_3\rangle
\, .\ee
Our templates will be a set of discretized partonic states corresponding to given points in phase space.
We wish to generate an ample number of template states to cover both two- and three-particle phase space for Higgs decay. To this end, we use a sampling Monte Carlo routine to generate points in a region of phase-space of final-state momenta, with a density proportional to the differential cross section, as described in Sec.\ \ref{meopt} above.

We assume that the lifetime of the Higgs particle is long enough so that its decay is incoherent with the evolution of
the hard scattering.
Thus, the rest frame of the particle is identical to the CM frame of
its decay products. Starting from the hard parton-level matrix
elements, we can readily  find the distribution of template momenta in
the Higgs jet rest frame, and then, by a straightforward Lorentz
transformation, boost it to the lab frame.

 For the two-body Higgs decay, two angles define
the two-body state of the daughter particles.  We choose these
as the polar and azimuthal angles in the Higgs rest frame, relative to the boost axis that
links the Higgs rest frame with the lab frame.
In these coordinates, the Lorentz invariant differential cross section for the two-body Higgs decay has the simple form
\be
\frac{d\Gamma}{\Gamma_0} = \frac{1}{4 \pi}\, d\phi \,d\cos\theta,\label{2_body_ps}
\ee
where $\Gamma_0$ is the tree-level decay width for a light Higgs,
\be
\Gamma_0 = \frac{N_c G_F m_H m_{b}^2}{4 \sqrt{2}\pi} \left(1- \frac{4 m_b^2}{m_H^2}\right)^{3/2},
\ee
with $m_b$ the $b$-quark mass, $m_H$ the Higgs mass, $G_F$ the Fermi coupling constant and $N_c=3$ the color factor.

By straightforward Lorentz transformations of particle momenta, the
two angles identified above determine the energies and directions of
the two decay products of the Higgs at LO. 
The two angular parameters $\theta$ and $\phi$ are distributed
according to (\ref{2_body_ps}). 
We generate a large
set of template states so that we are confident of identifying the
peak value of overlap. We can now encode the two physical angles in terms of row and column numbers, corresponding to the data discretization scheme. Each template consists of the information ($\rm row_a, column_a$, $E_{\rm a}$) for each of the two daughter particles.  We exclude those templates having polar
angles (in the lab frame) larger than the cone size $R$. 

Next, we consider the templates motivated by the on-shell
next-to-leading order decays of the Higgs. In this case, more
variables enter the problem.   They can be chosen as
$\{x_1,x_2,x_3\}$, $x_i$ denoting the fractional energies of the
(massless) partons in the jet CM frame (normalized so that $0<x_i<1,\sum x_i
=2$), and three Euler angles $\psi, \theta,\phi$ to parameterize the orientation of the decay products in the Higgs rest frame (see Appendix A). 
Since the Higgs boson is a scalar, its decay products are spherically
symmetric and therefore distributed uniformly in $\psi$, $\phi$ and
$\cos\theta$.  Specifically, in the rest frame of the scalar Higgs,
the $b$, $\bar b$ quarks and the gluon, $g$ have energy
fractions given by $x_1$, $x_2$ and $x_3$, respectively, distributed
according to the differential cross section,
\be
\frac{d\Gamma}{\Gamma_0} = \frac{1}{8 \pi^2} C_F \alpha_s  \frac{(1-x_1 -x_2)^2+1}{(1-x_1)(1-x_2)}\, dx_1 dx_2 \,d\psi \,d\cos\theta \,d\phi.  \label{3_body_ps}
\ee
Here $\alpha_s$ is the coupling constant and $C_F= (N^2-1)/2N $. 
There exists a certain region of phase space, corresponding to the soft or collinear emission of a gluon, in which the three-parton templates simulates a two-parton template. This is precisely the region in which the differential distribution (\ref{3_body_ps}) contains divergences, which cancel in the total transition probability after the inclusion of virtual gluon exchange diagrams. We regulate these divergences in terms of the invariant mass of the pair of nearly-collinear partons, which has a simple relation  to the variables $x_i$.     For a partonic configuration to correspond to a three-particle template, we require
\be
y_{ij}=\frac{(p_i+p_j)^2}{m_H^2}=(x_i+x_j-1) >y ,\,  \label{invariant_mass_cut}
\ee
where $y$ is a cut-off parameter.  In the following, we take $y=0.05$\@.

As for the two-parton templates, we generate a large number of three-parton templates, of order one million, distributed according to (\ref{3_body_ps}), while satisfying the constraint (\ref{invariant_mass_cut}).
By Lorentz transformations of particle momenta, analogous to the
two-particle case, the two energy fractions and the three angles
identified above determine the energies and directions of the three
decay products of the Higgs at NLO. As for the discretization of the
data, we again encode two physical angles in terms of row and column
number corresponding to the data discretization scheme. A given
three-particle template consists of a list ($\rm row_a, column_a$,
$E_{\rm a}$, a=1,2,3) for each of three daughter particles of NLO
Higgs decay ($b$, $\bar b$ and $g$). We also exclude those templates having particles whose polar angles in the boosted frame,
relative to the jet axis, are larger than the cone size $R$. 

Clearly, templates cannot be meaningful for zero momentum partons.
 However,  the invariant-mass cut Eq. (\ref{invariant_mass_cut}) and the requirement that the templates are within a cone of size $R$ imposes  a lower cutoff on the energies of the template partons,
\be
\frac{2 \,m_H^2 \, y}{P_0 \,R^2} \lesssim E_{\rm a}\, . \label{RcutE}
\ee
For the benchmark parameters used here, $P_0=1000\GeV$,
$m_{H}=120\GeV$, $y=0.05$, the corresponding cutoffs are $E_{\rm a}
\gtrsim 3(9)\GeV$ for $R=0.7 (0.4)$. Although the cutoff corresponding
to $R=0.7$ happens to be particularly close to the hadronization
scale, we checked that imposing a more stringent, explicit cut on our templates, namely $E_{\rm a} \geq 20\GeV$, did not have a sensible effect on the results.

\subsection{Two- and three-particle template overlap}\label{23bodytemplates}

We are now ready to implement Eq.\ (\ref{overlap2}) for the Higgs, by defining an overlap between templates, $|f\rangle $, and
jet states, $|j\rangle$,  $Ov=\langle j | f \rangle$.
Defined as above, our two-parton templates each have two cells corresponding to two daughter partons ($q$ and $\bar q$) with their row and column numbers determined by the data discretization scheme. In addition, we have three-parton templates each having three cells corresponding to three daughter partons ($q$, $\bar q$ and $g$) with their row and column numbers determined by the same discretization scheme.

We compute the overlap between data state $j$ and two- or three-body template $f$
from the unweighted sum of the energy in the nine cells 
of state $j$ surrounding and including the occupied cells of template state $f$,
\be
Ov_N(j,f) =
{ \rm{max}}_{\tau^{(R)}_N}\ 
\exp\left[\, -\ \sum_{a=1}^N \frac{1}{2\sigma^2_a}\left(  
\,  \sum_{k=i_a-1}^{i_a+1}\sum_{l=j_a-1}^{j_a+1} E{(k,l)}   - E(i_a,j_a)^{(f)}\, \right)^2
   \right]\, ,
 \label{Nparticletemplate}
 \ee
where $N=2$ or $3$\@. Here, $E(i_a,j_a)^{(f)}$ is the energy in the template state for  
particle $a$ whose direction is labelled by indices $i_a$ and $j_a$, according to the discretization table
described in Sec.~\ref{event_discretization}\@.   If one of the sums extends outside
the jet cone, we set the corresponding energies $E(k,l)$ to zero.
We fix $\sigma_a$ (for the $a$th parton) by that parton's energy,
$\sigma_a=E(i_a,j_a)^{(f)}/2$.

The use of $N=2$ and $N=3$ templates follows three steps. Starting
from a hard massive jet on angular scale $R$, one identifies the two
subjets within it using two-parton templates. Within the same angular
region, one then further takes the three hardest subjets that appear
using three-parton templates, if the energies of three subjets satisfy
Eq. (\ref{RcutE})\@. If this condition is not satisfied, only the
two-body templates are used. The whole procedure is illustrated in Fig.~\ref{procedure}. 

\begin{figure}[hptb]
\begin{center}

\begin{tabular}{c}
\includegraphics[width=.9\hsize]{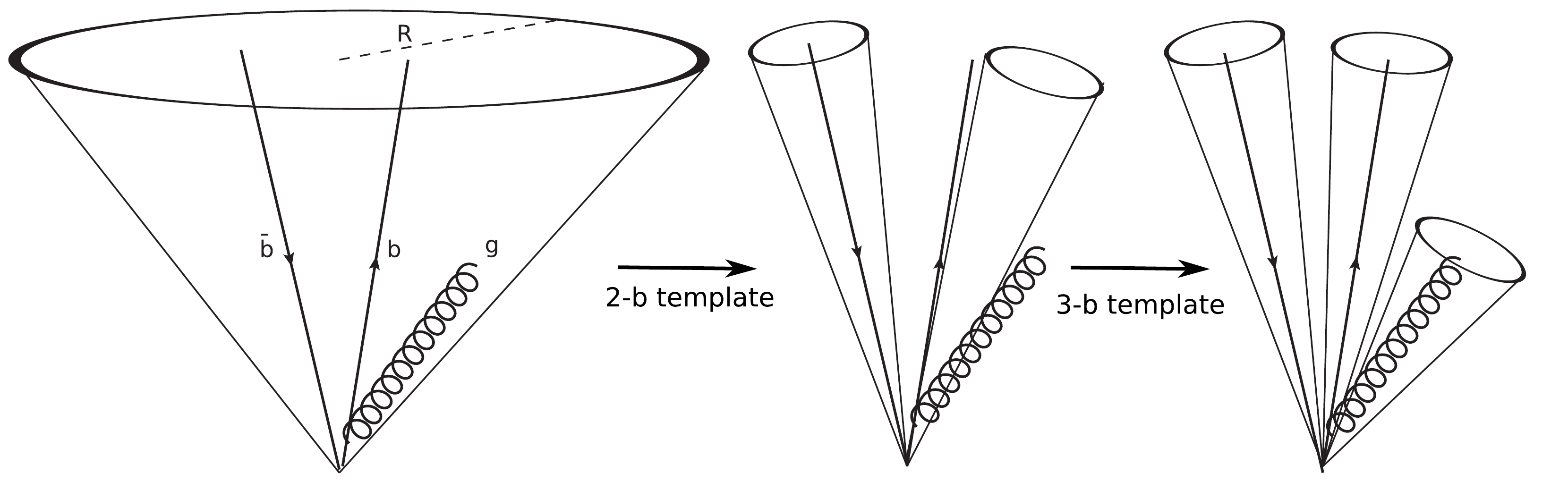}
\end{tabular}
\end{center}

\caption{Stages of the template method described in text.}
\label{procedure}
\end{figure}

\subsection{Parton-level studies}
\label{sec:partonlevel}

To check the consistency of our method, we first test our ideas using
parton-level events, which are simulated by a Monte Carlo
sampling of Eq. (\ref{3_body_ps}). The Monte Carlo samples used for this validation include final state
radiation up to ${\cal O}(\alpha_s)$. The events are conveniently
regularized as in Eq. (\ref{invariant_mass_cut}) via an invariant-mass cut algorithm with $y =0.05$\@. The signal samples thus contain only $q$ and $\bar q$ quarks and up to one gluon $g$ in the final state. These events are not run through full parton showering and hadronization simulations. Consequently, the generated partons in each event are distributed according to the same squared matrix elements used  in the template generation, Eqs. (\ref{2_body_ps}) and (\ref{3_body_ps}).

In Fig. \ref{validate}, the results are shown for the two- and
three-body overlap given by Eq. (\ref{Nparticletemplate}).
The data shows that, as required for consistency, each peak value $Ov_2$ ($Ov_3$) is close to unity
for all the events in our Higgs decaying into a $b\bar b$ ($b\bar b
g$) sample, around 80\% (20\%) of the total (see Fig.~\ref{jade_br} in
Appendix~B). 
Of the 22\% with $Ov_3>0.8$, 14\% have small $Ov_2$ while 8\% have
substantial two-body overlap, $0.2<Ov_2<1$\@. 
As we will see below, the effect of showering is to spread out Higgs
decays over the full range of $Ov_2$ and $Ov_3$\@.

\begin{figure}[hptb]
\begin{center}

\begin{tabular}{c}

\includegraphics[width=.35\hsize]{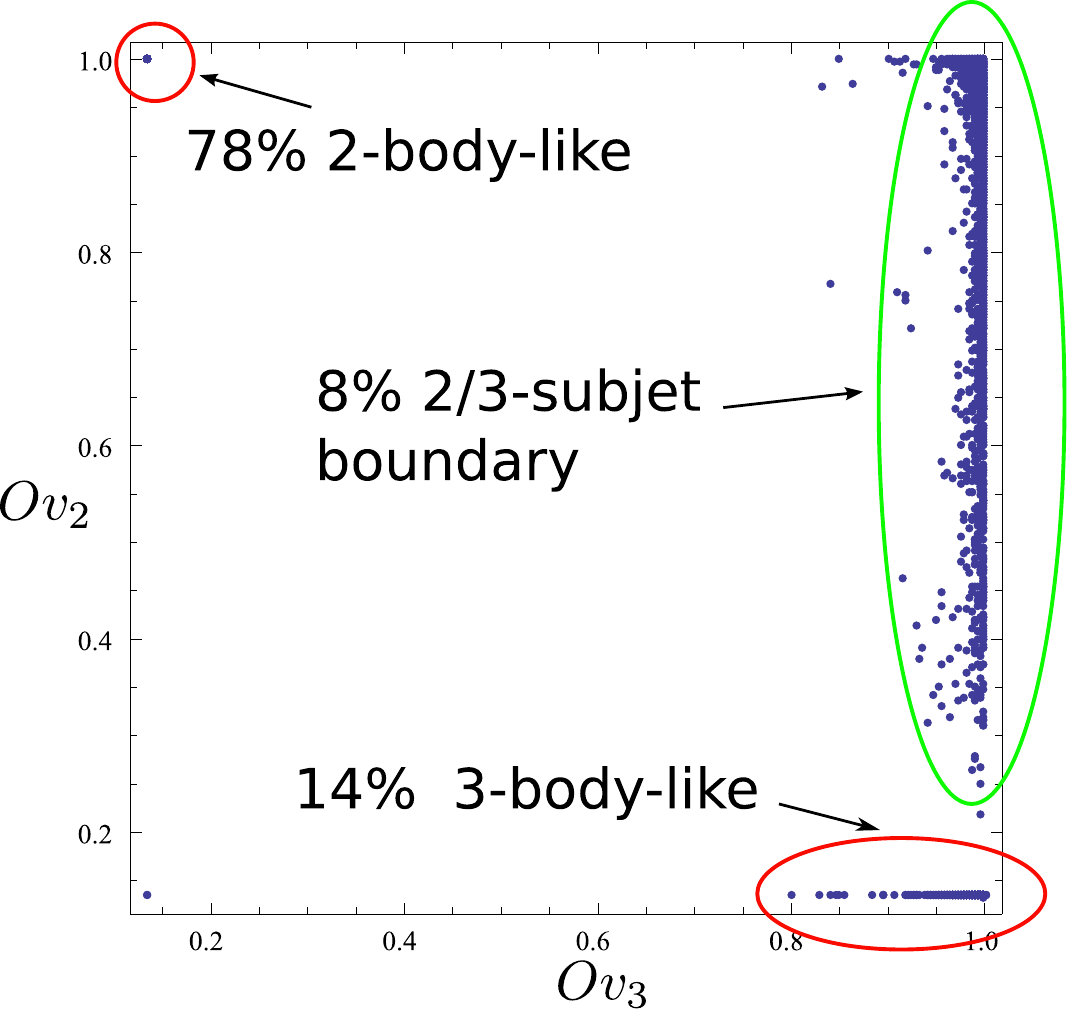}

\end{tabular}
\end{center}

\caption{A scatter plot of two-parton template overlap {\it vs.} three-parton template overlap  for LO parton-level MC output for Higgs decay, with jet energy, $P_0=1\TeV$, $m_{H}=120\GeV.$ }\label{validate}
\end{figure}

\subsection{Template overlaps for Higgs and QCD jets} \label{overlap_plots}

We now proceed to 
test the ability of two- and three-parton templates
to distinguish between energetic Higgs jets from QCD jets. We apply templates to events generated by {\sc Pythia}\cite{pythia8}, {\sc MG/ME}\cite{madgraph}, and {\sc Sherpa}\cite{sherpa}. The event selection was described in Sec. \ref{event_discretization}.

In Fig. \ref{higgs_bin} we exhibit typical overlap distributions for showered Higgs and QCD jets (for the same jet mass and energy) for event generators {\sc Pythia} (version 8) for $2\to 2 $ process without matching, {\sc MadGraph/MadEvent (MG/ME)} (with MLM matching interfaced into {\sc Pythia V6.4}), and {\sc Sherpa} 1.2.1 (with CKKW matching). 
In the plots on the left panels of Fig. \ref{higgs_bin}, we compare
the template overlap distribution from Eq. (\ref{Nparticletemplate})
with $N=2$ for Higgs and QCD jets. The corresponding plots on the
right panels of Fig. \ref{higgs_bin} show similar distributions when
three-particle templates ($N=3$) are used.  Clearly, showering smears the Higgs distributions significantly, although Higgs
events are concentrated at larger peak overlaps than QCD events for both $N=2$ and
$N=3$\@. These differences, although not adequate to isolate a small
signal, can serve as a foundation for further enrichment by use of
additional analysis of template and jet states, as discussed
below.\footnote{Note also the moderate variations between the different Monte Carlo generators. Since the template overlap method depends on the precise radiation pattern within a jet, these variations are expected.}

\begin{figure}[hptb]
\begin{tabular}{cc}
\includegraphics[width=.48\hsize]{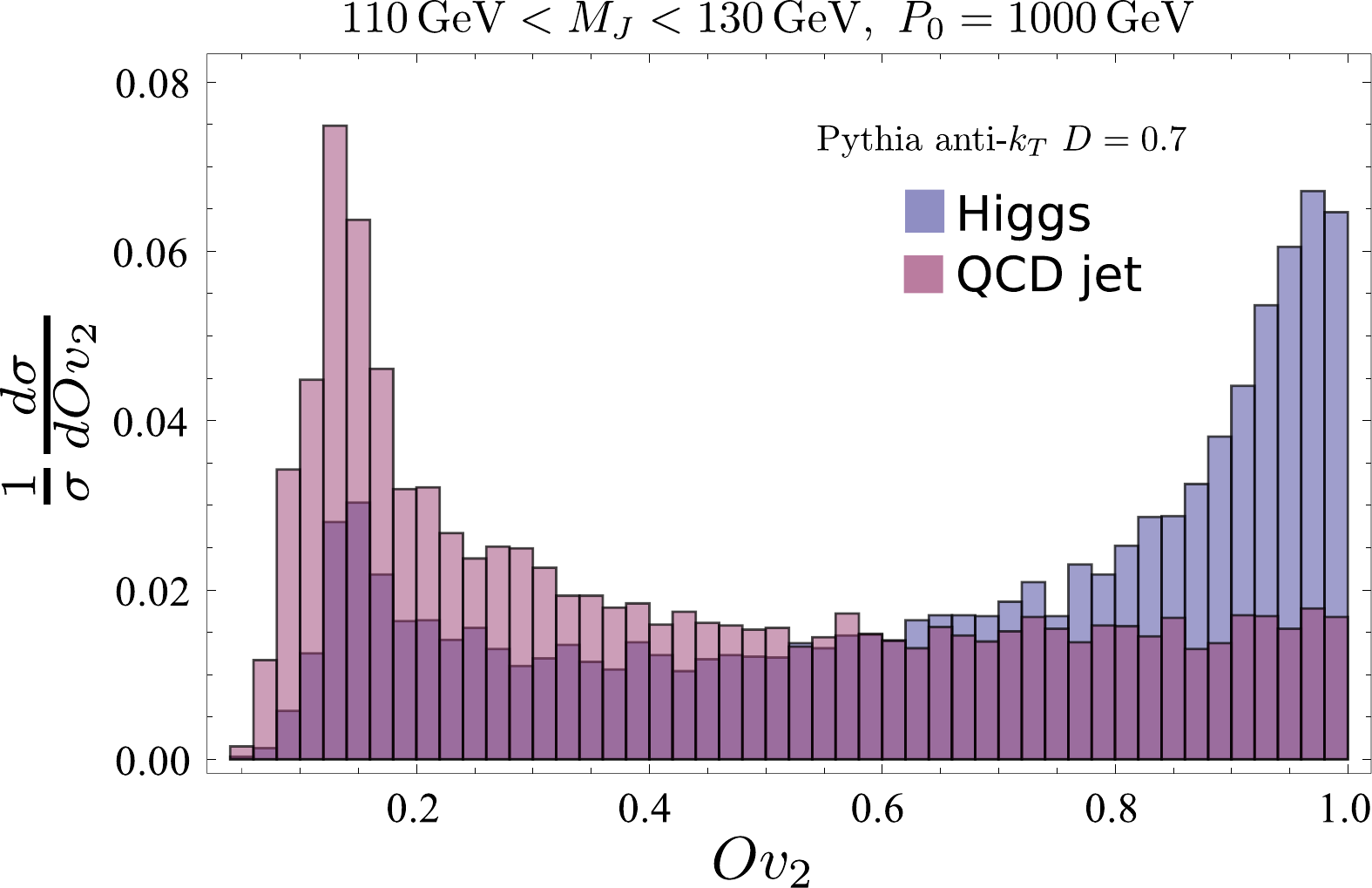}
\includegraphics[width=.48\hsize]{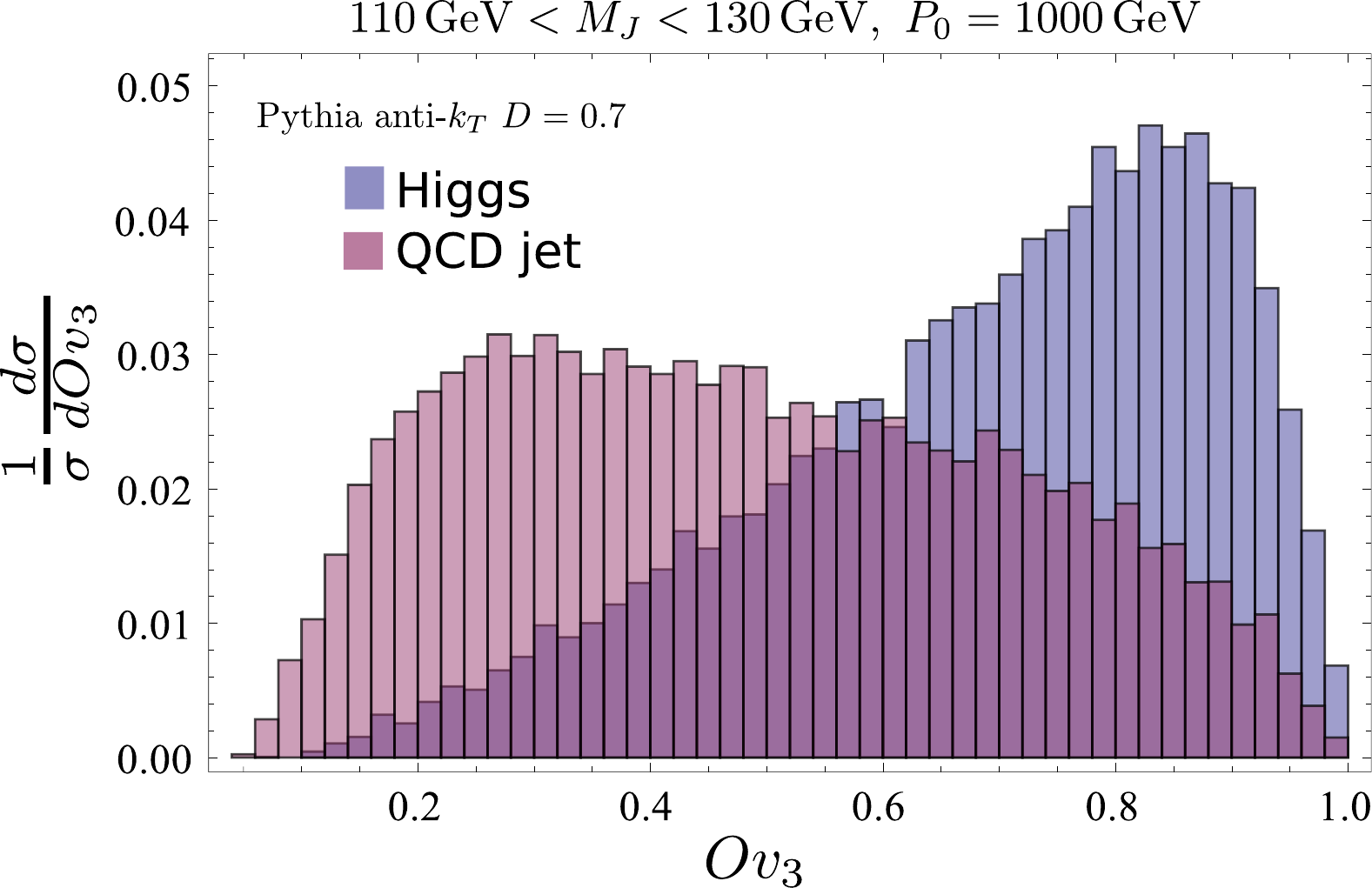} \\
\includegraphics[width=.48\hsize]{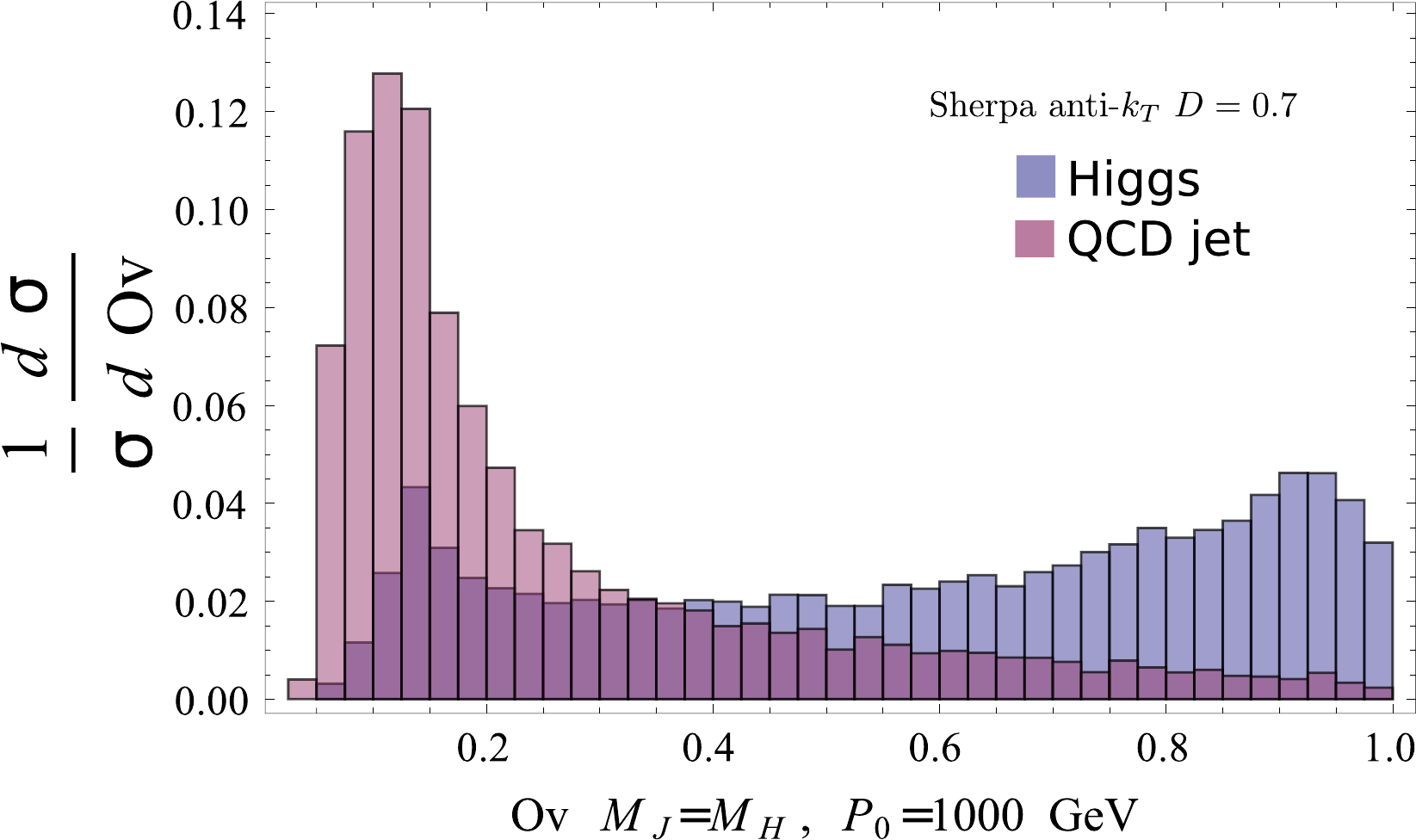}
\includegraphics[width=.48\hsize]{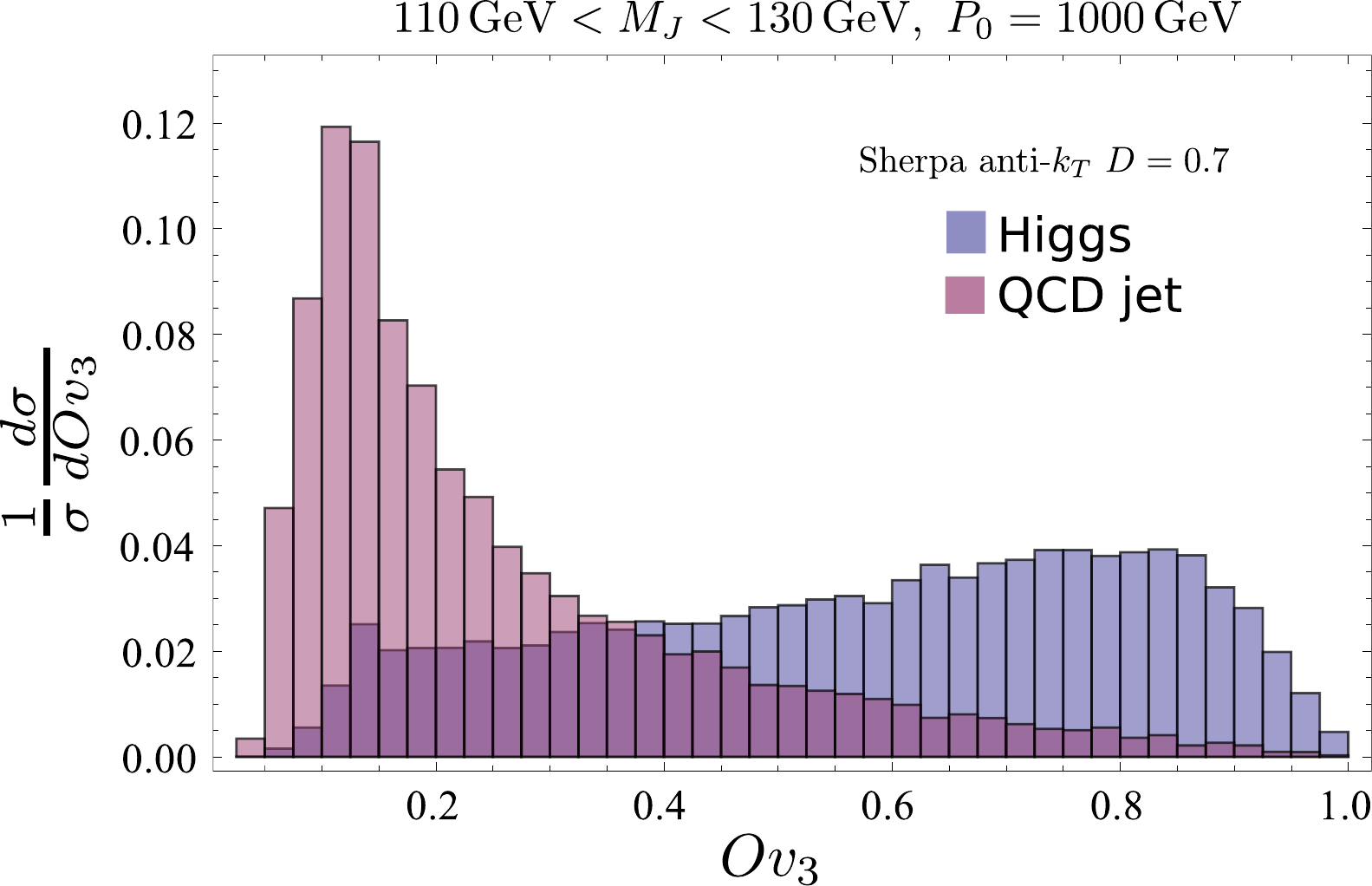}\\
\includegraphics[width=.48\hsize]{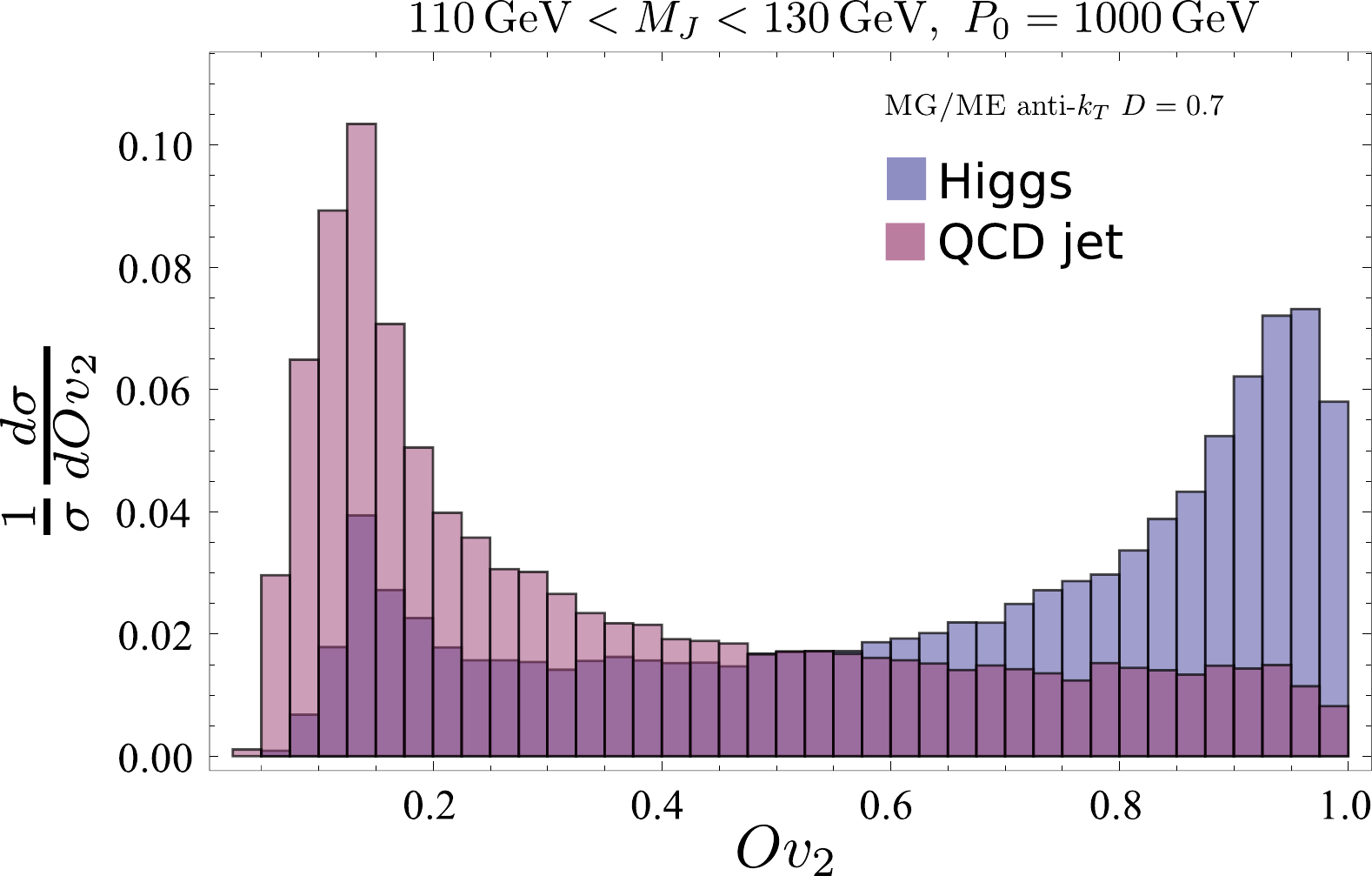}
\includegraphics[width=.48\hsize]{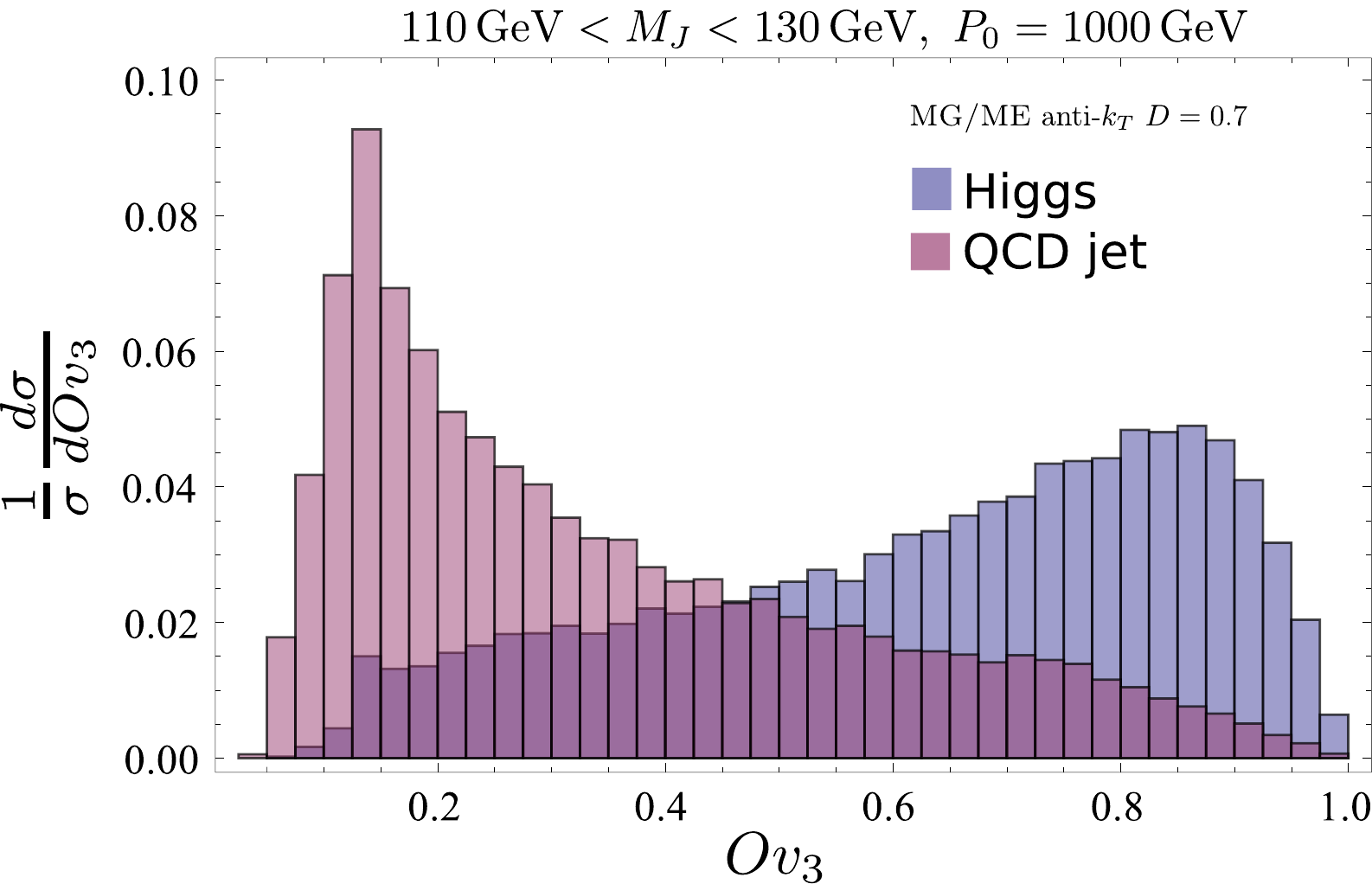}
\end{tabular}
\caption{
Comparison of histograms of template overlap $Ov$ with Higgs jets and QCD jets from different MCs [from top to bottom, {\sc Pythia}, {\sc Sherpa} and {\sc Madgraph}], for $R=0.7$,  950 GeV$\le P_{0} \le$1050 GeV, 110 GeV$\le m_J \le$130 GeV and  $m_{H}=120$ GeV using 2-body templates (left) and 3-body templates  (right).
}\label{higgs_bin}
\end{figure}

\section{Single-variable discriminants}\label{Sec:Single_variable_discriminants}

While overlap distributions help enrich samples of Higgs events by capturing the 
constrast 
 in the radiation pattern for the Higgs signal and QCD background, there is still a large tail of QCD background events with relatively large $Ov$. 
 We proceed to develop and study  several single-variable
 discriminants, which provide us with additional tools to maximize the
 separation between QCD and Higgs jets. Many of these variables,
 although intimately related, are worth studying individually, since they can be used to highlight different characteristics of jet substructure.

Infrared-safe observables, such as jet shapes, guarantee that we can
make meaningful comparisons between theoretical computations and their
analogous experimental measurements. The method identifies the
template that a physical state most closely resembles.  We can also
analyze, for example, angular distributions of template partons,
considered as subjets, without the energy weights in event shapes. The
template can thus be used to identify jet substructure~\cite{Butterworth:2008iy,Kaplan:2008ie,Kribs:2009yh,Chekanov:2010vc,arXiv:1010.5253,arXiv:1011.4523,arXiv:1006.3213,arXiv:1010.0676,arXiv:1008.2202,arXiv:1012.2077,arXiv:1010.3698,  arXiv:1102.0557,arXiv:1007.2221,arXiv:1104.1646,arXiv:1011.1493,Pruning,Krohn:2009th,Soper:2010xk,arXiv:1102.3480}. The overlap method enables us to make subjet identification by providing a mapping between energy-unweighted 
 variables and the template that defines the energy flow distributions. In what follows, we describe these two categories of tools that will allow us to further reject background events with a large $Ov$.

\subsection{Planar flow}\label{Sec:jet_shapes}

Jet shape variables are an especially interesting class of observables for jet studies, and have received considerable attention in the past several years in the context  of boosted jet identification \cite{Thaler:2008ju,Almeida:2008yp,arXiv:1101.2905}.  The common feature of all jet shapes is that they involve moments of the energy of observed particles  and are thus  smooth functionals of energy flow within a jet.  In this manner, they are complementary to the information provided by template overlaps, which is associated with jumps and spikes in energy flow. 

Following Ref.\ \cite{Almeida:2010pa}, we will make use of the jet
shape planar flow in the form,
\be
Pf={4\,{\rm det}(I_{\omega})\over{\rm tr}(I_{\omega})^2},
\label{Pfdef}
\ee
where $I_{\omega}$ is defined by,
\be
I^{kl}_{\omega}= {1\over m_J} \sum_i {\omega}_i \frac{p_{i,k}}{{\omega}_i}\,\frac{p_{i,l}}{{\omega}_i}\, ,
\ee
with $m_J$ the jet mass, ${\omega}_i$ the energy of particle $i$ in the jet,
and $p_{i,k}$ the $k^{th}$ component of its transverse momentum relative to the 
axis of the jet's momentum.
Jets attributed to two-body final states have a differential jet function fixed at zero planar flow,
\be
\frac{1}{J}\left(\frac{dJ}{dPf}\right)_{\rm 2 \; body} =\delta(Pf).
\ee
This would apply 
at leading order for events with highly boosted Higgs and QCD jets. On the other hand realistic QCD and Higgs jets have nonzero $Pf$, because of QCD radiation effects that smear the distribution. 

We expect soft radiation from the boosted color singlet Higgs to be concentrated between the $b$ and $\bar b$ decay products.   This is to be contrasted to a jet initiated by a light parton, whose color is correlated with particles in other parts of the
 event, producing radiation in the gaps between those particles and the jet system.
Therefore, we expect that planar flow for Higgs jets will be peaked toward a lower value than that of QCD jets. In the studies we show below, the combination of $Ov$ and $Pf$ gives a
strong background (QCD) suppression with quite substantial signal (Higgs decay) efficiency.

\begin{figure}[hptb]
\begin{center}
\begin{tabular}{cc}
\includegraphics[width=.9\hsize]{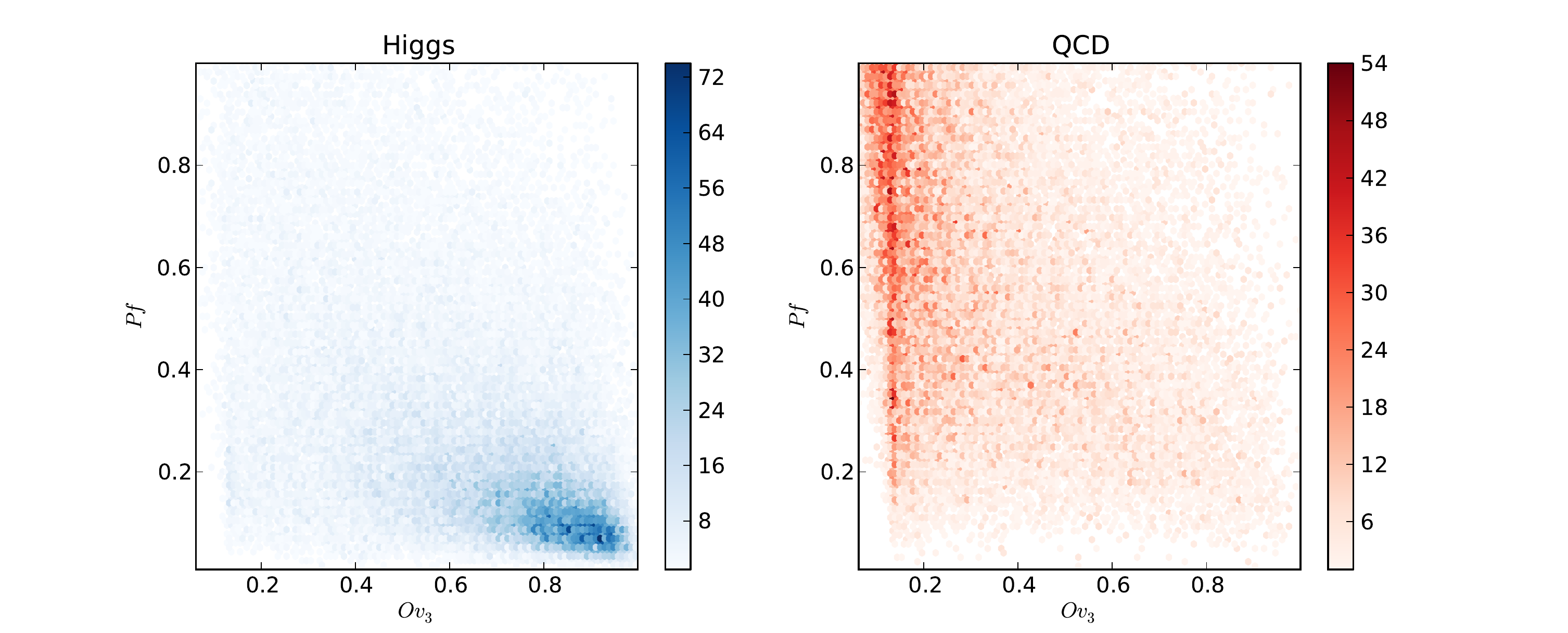}
\end{tabular}
\caption{
Density plots of planar flow $Pf$  {\it vs.}\ template overlap $Ov$ for Higgs jets  and QCD jets  from  {\sc Pythia}\cite{pythia8}, for $R=0.7$,  950 GeV$\le P_{0} \le$1050 GeV, 110 GeV$\le m_J \le$130 GeV using three-body templates.
}\label{higgs_probe_function_pf_comparison}
\end{center}
\end{figure}

 Figure \ref{higgs_probe_function_pf_comparison} shows the two-dimensional distributions of MC events (obtained via {\sc Pythia  \cite{pythia8}}) in the $Pf$ vs. $Ov_3$ plane for the signal and background. The scatter plot shows that signal events cluster around large overlap while, at the same time, $Pf$ is essentially below 0.2.  By contrast, QCD events tend to be spread over the entire area.  These plots also confirm our expectation that Higgs jets tend to have smaller $Pf$ values than QCD jet events (for the same ratio $m_J/P_0$). Clearly, any set of events chosen from the bottom right of these plots, with $Pf<Ov_3$, is highly 
 enriched  
 in three-body Higgs events compared with background. The clear difference in these scatter plots shows the potential of the template overlap method.

\subsection{Template variables}\label{Sec:Template_variables_methods} 
 
We will refer to the second set of variables for discriminating jets
as template variables. 
We focus on variables which are sensitive to differences between jet
radiation  for signal and background. For the particular case of the Higgs boson signal, we aim to show the potential discriminating power of this category of variables.    In future work it should be possible  to improve upon our na\"{i}ve analysis below, which is based on simple rectangular cuts.  

We can test these ideas on planar flow. The $Pf$ computed from the
template of any event is a template variable, and can be compared to
the corresponding $Pf$ of the event. For comparison purposes, in
Fig. \ref{validate_pf} we show the two-dimensional distributions for
the actual planar flow and its template variable version for Higgs
jets. The data corresponds to Higgs jets from parton-level MC output
based on Eq. (\ref{3_body_ps}) (left) and {\sc Pythia} (right).  The parton level plot validates the
use of templates, since points are concentrated on the diagonal. The
distribution from {\sc Pythia} is slightly different, but the
correlation between the actual planar flow and its template version is
evident. One can see that the effect of showering is to smear the $Pf$
distribution to larger values.
\begin{figure}[hptb]
\begin{center}
\begin{tabular}{cc}
\includegraphics[width=.9\hsize]{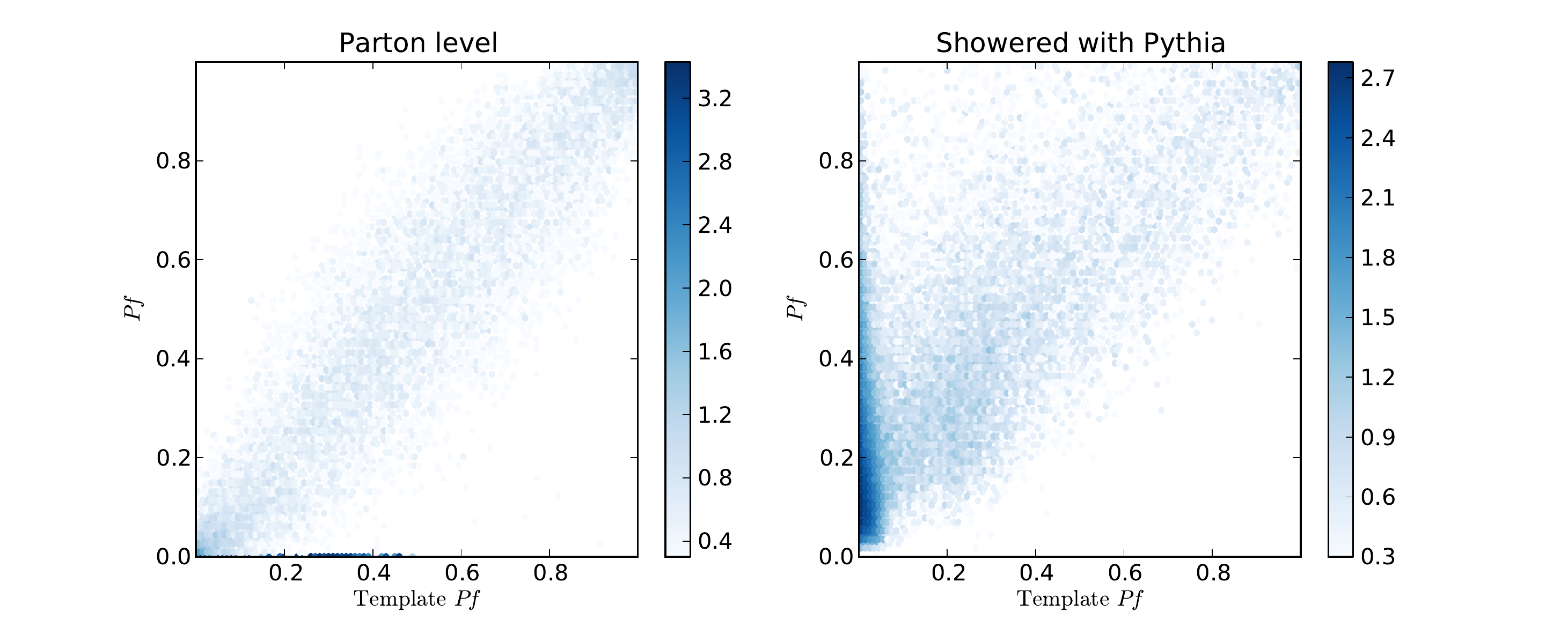}
\end{tabular}
\end{center}
\caption{Density plots of   $Pf$ {\it vs.} partonic template $Pf$ for Higgs decay events from LO parton-level MC output  (left) and {\sc Pythia} (right), with $P_0=1\TeV$, $m_{H}=120\GeV.$ The intensity of the shading is proportional to the density of points.}\label{validate_pf}
\end{figure}

We first consider variables constructed out of templates with the minimum number of particles. This has been discussed in detail in~\cite{Almeida:2010pa}, so our discussion will be brief. As described in~\cite{Almeida:2008yp}, at lowest order the signal phase space for the Higgs decay is characterized by simple
kinematic parameters.   For example, 
in 
Higgs
decays, there are only two variables that characterize the decay $h\rightarrow b\bar b$. 
One of them is the jet energy $P_0$, and a convenient choice for the other variable is the angle, $\theta_s$, between
the 
jet axis and the softer of the two particles.
At fixed $m_J/P_0$, Higgs events tend to be peaked towards smaller
values of $\theta_s$ than QCD jets~\cite{Almeida:2008yp}. Therefore,
even within the two-body description, $\theta_s$ already provides
useful information of jet energy flow. An analysis of this template
variable and its application to two-body Higgs decay was presented in~\cite{Almeida:2010pa}.

Once we include the contribution from leading order radiation to the Higgs decay, 
the phase space is now characterized by five variables, $(x_1,x_2,\psi,\theta,\phi)$ as defined 
in Sec. \ref{construction}. This gives us more freedom to contrive
template variables. In the following, we discuss distributions in
these variables, beginning with the energy fractions, $x_1$ and
$x_2$\@. Here and below, $x_1$ is the energy fraction for the
$b$~quark while $x_2$ is that of $\bar{b}$ in the template.

We may anticipate that $x_1$ and $x_2$ are not ideal discriminants
between Higgs decay and QCD events, because both distributions have
singularities at $x_1=1$ and $x_2=1$, which reflect collinear and soft
gluon emission. This is seen in Fig. \ref{densityx1x2}, where we show two-dimensional distributions for $\{x_1,x_2\}$ for Higgs and QCD peak templates. 
The top panels show the $\{x_1,x_2\}$ distributions at the {\sc Madgraph} level before showering, 
and the bottom panels after events are showered and jets reconstructed with {\sc Pythia} and 
{\sc Fastjet}. We have also imposed a quality cut on the three-body template overlap $Ov_3>0.6$. 
We can see that the discriminating power of the energy fractions is reduced by requiring an overlap
 cut in the showered events.  While one might still be able to improve signal to 
 background by drawing a contour around the collinear regions, it is clear that the $\{x_1,x_2\}$ 
 variables alone will not be very effective in discriminating boosted Higgs jets from QCD background,
  and in any case, in these regions we expect resummation effects to be important.

\begin{figure}[hptb]
\begin{center}
\begin{tabular}{cc}
\includegraphics[width=.9\hsize]{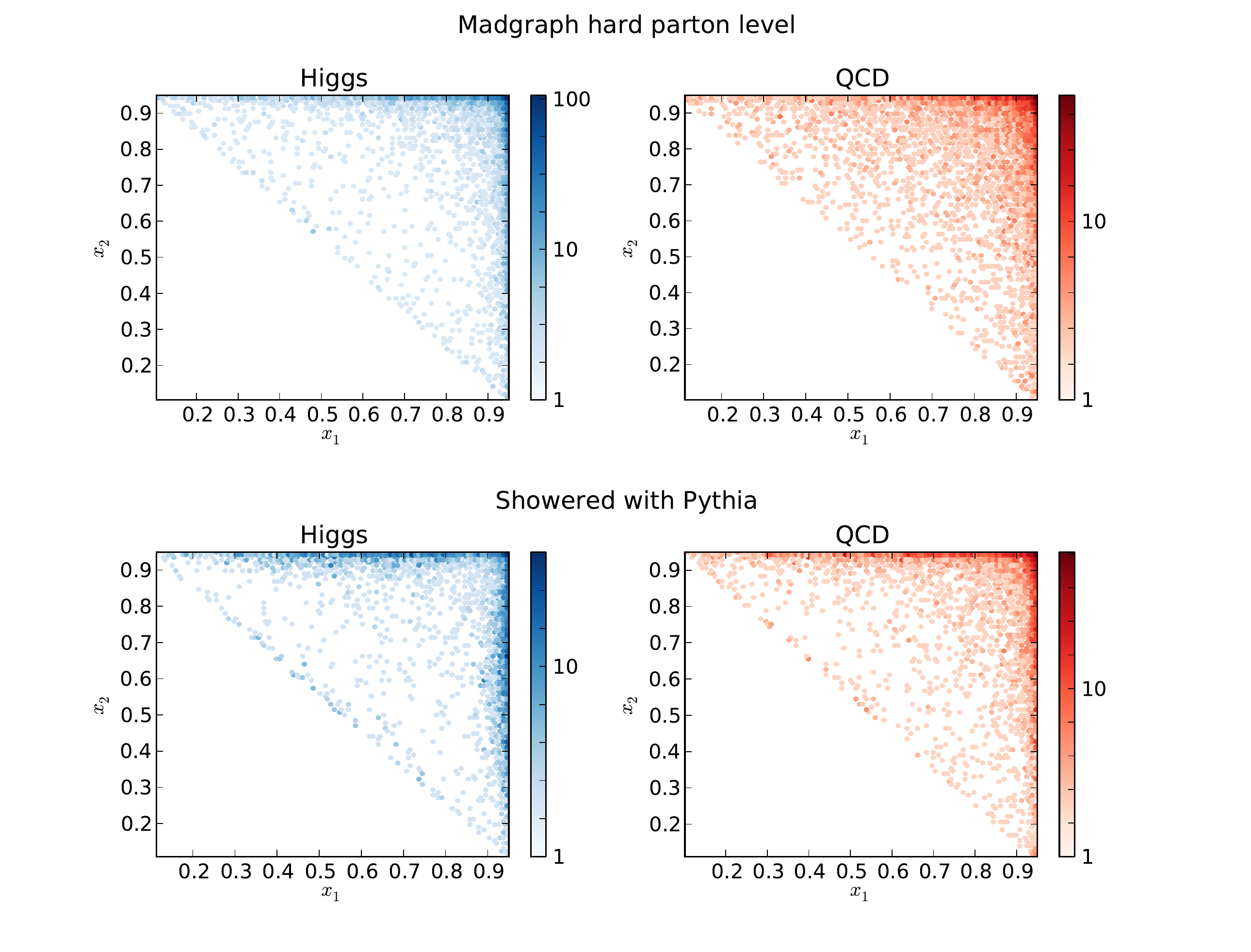}
\end{tabular}
\end{center}

\caption{Density plots of $x_1$ vs. $x_2$ for Higgs signal (blue) and QCD background (red), for the LHC.  {\sc Madgraph} hard parton-level (top) and showered jet level with {\sc Pythia } (bottom).  Both are shown only in the Higgs mass window $110 \GeV < m_J< 130\GeV$, with $P_0=1\TeV$, $m_{H}=120\GeV.$} \label{densityx1x2}
\end{figure}

Template variables based on angular distributions are more promising
than energy fractions. To be specific, we will consider the following
(lab frame) variables, and look at their distributions. 

\begin{itemize}
\item The angles between the jet axis and the template momenta $\theta_{iJ}$,
\be
1-\cos\theta_{iJ} =  \frac{z\,x_i\,m_J}{2 E_i} \ , \label{thetai}
\ee
with $z=m_J/P_0$\@. 
\item The angular separations: $\theta_{12}$, $\theta_{13}$, $\theta_{23}$,
\be
1-\cos \theta_{ij} =  \frac{(x_i+x_j-1)\,m_J^2}{2 E_i E_j} \ . \label{thetaij}
\ee
\item The angle between the jet axis and the softest of the partons: $\tilde \theta_s$,
\be
1-\cos\tilde \theta_{s} =  \frac{z\,x_s\,m_J}{2 E_s} \ ,  \label{thetas}
\ee
where $E_s={\rm min}\{E_i\}$\@. 
\item $r_{\theta}= {\rm min}
  \{\theta_{13}/\theta_{12},\theta_{23}/\theta_{12} \}$, found by
  finding the minimum of $(1-\cos\theta_{i3})$, $i=1,2$, given by
\be
{\rm min} \left\{\frac{(1-x_2)E_2}{(1-x_3)E_3}\,
  ,\,\frac{(1-x_1)E_1}{(1-x_3)E_3}\right\} \label{R_param} \ .
\ee
\item The three-body angular variable $\bar \theta$,
\be
\bar \theta= \sum_i \sin \theta_{iJ} \label{thetabar} \ , 
\ee
with $\theta_{iJ}$ given by Eq. (\ref{thetai})\@. 
\end{itemize}
The expression for the energy $E_i$ of particle $i$ is fairly simple, and is given in Appendix A, Eqs.~(\ref{boostedE1}-\ref{boostedE3}).

 The distributions of the variables Eqs.~(\ref{thetaij})-(\ref{thetabar}) are shown in Fig.~\ref{angles}. All of these variables are shown for anti-$k_T$ $R=0.7$ jets. Since our focus is on the difference in the shapes of various observables, all of the kinematic distributions are normalized to unity. The angular variables $r_\theta$ and $\bar \theta$ offer the promise of reasonable discriminating power, because they are directly tied to physical features of the signal, as follows.
 
\begin{figure}[hptb]
\begin{center}
\begin{tabular}{ccc}
\includegraphics[width=.33\hsize]{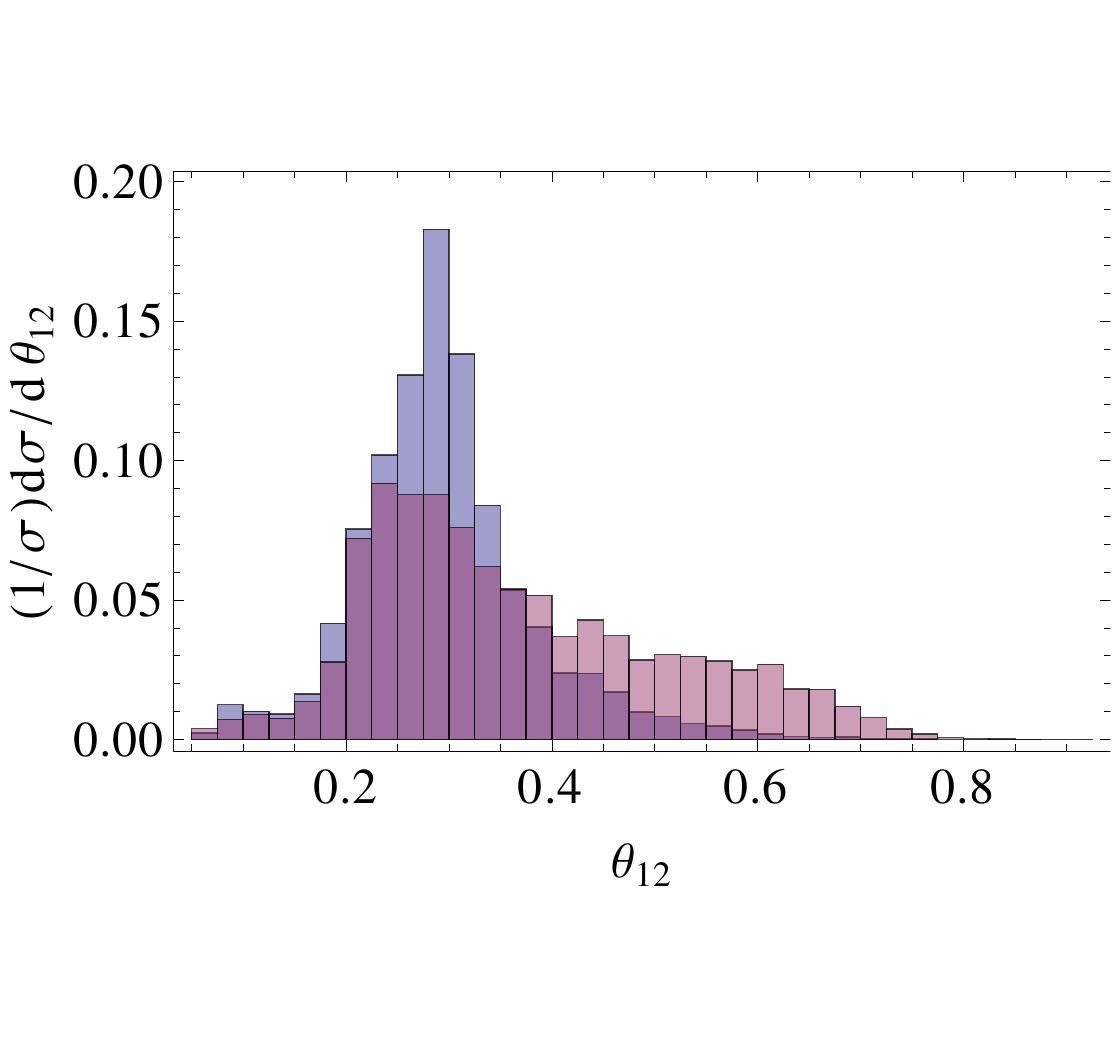}
\includegraphics[width=.33\hsize]{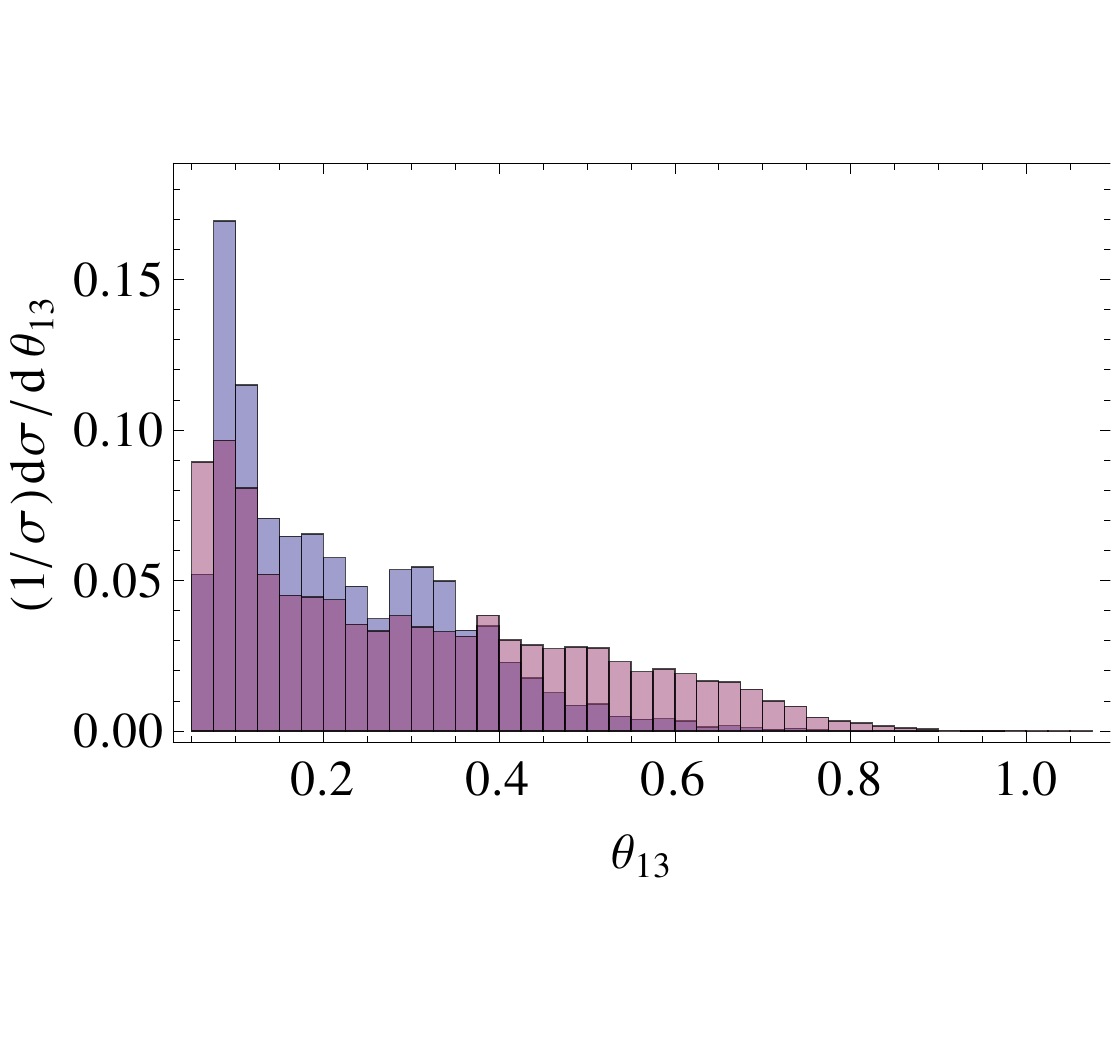}
\includegraphics[width=.33\hsize]{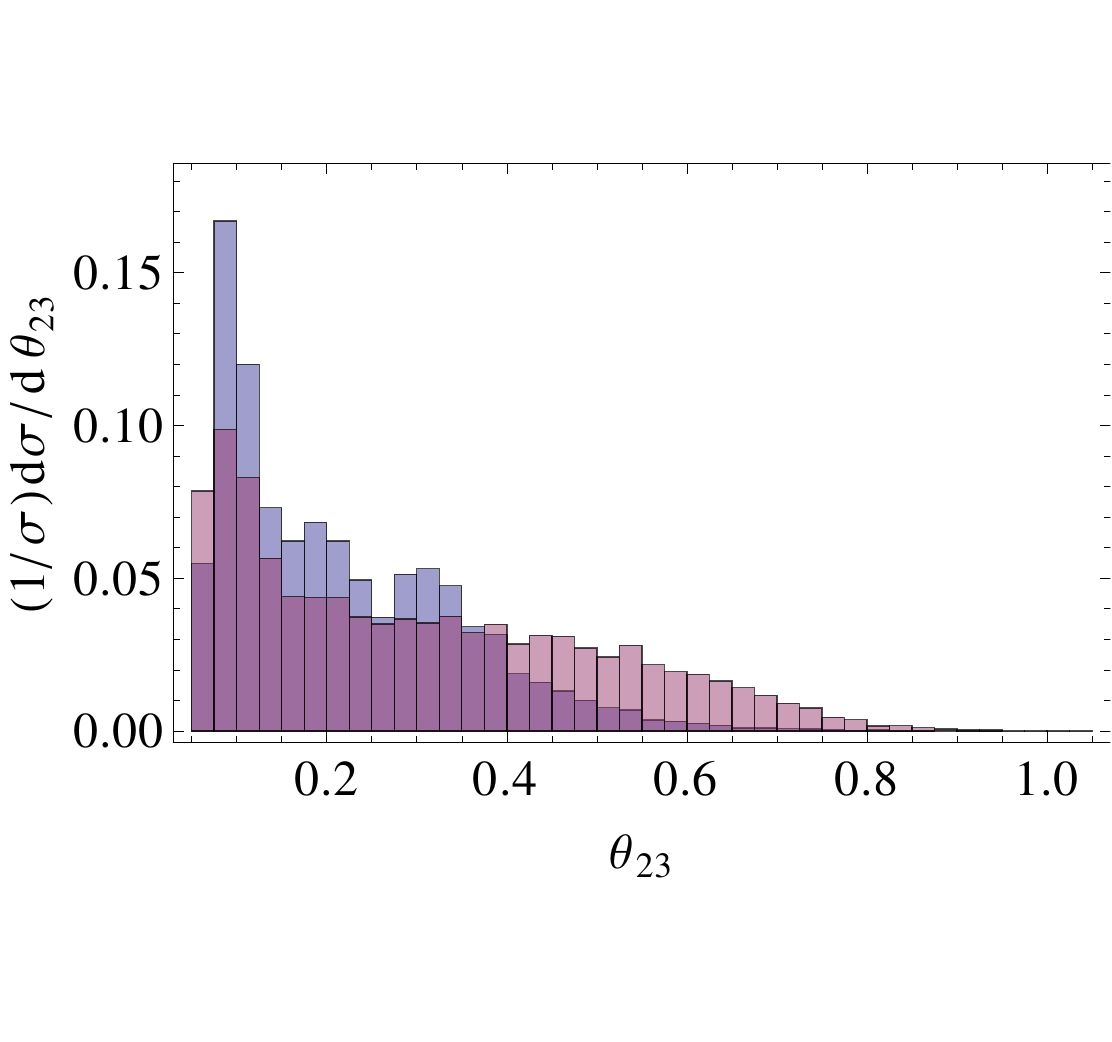}\\
\includegraphics[width=.33\hsize]{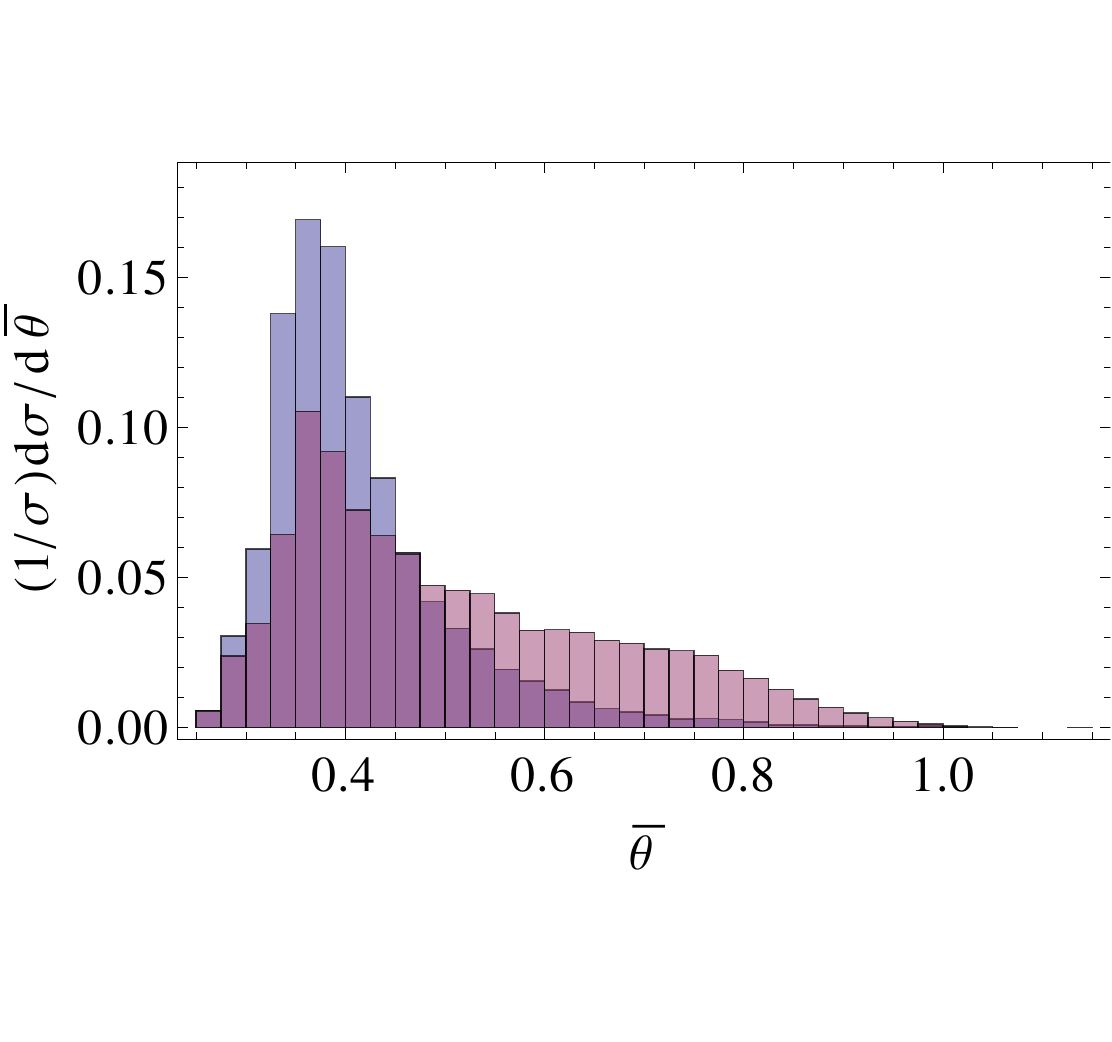}
\includegraphics[width=.33\hsize]{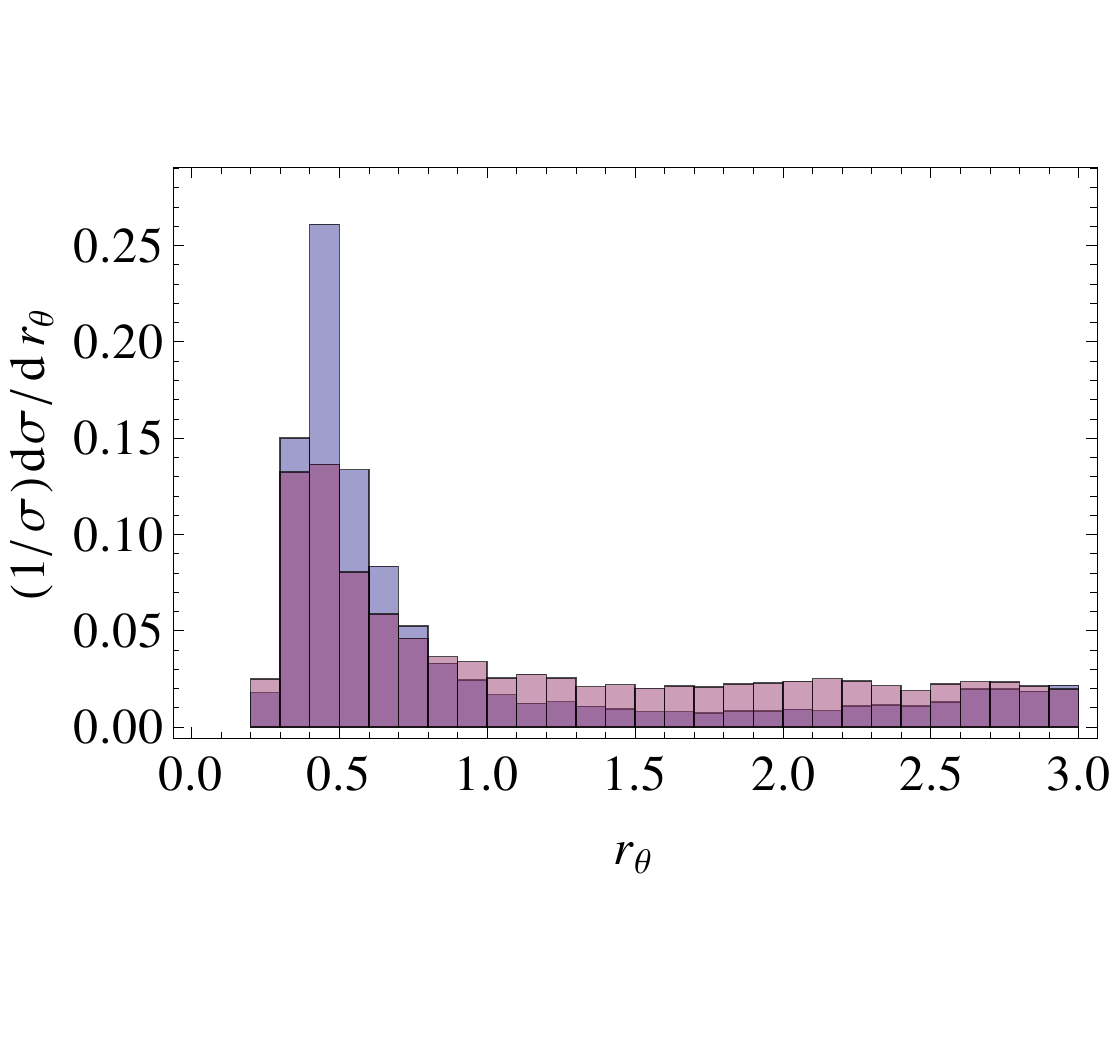}
\includegraphics[width=.33\hsize]{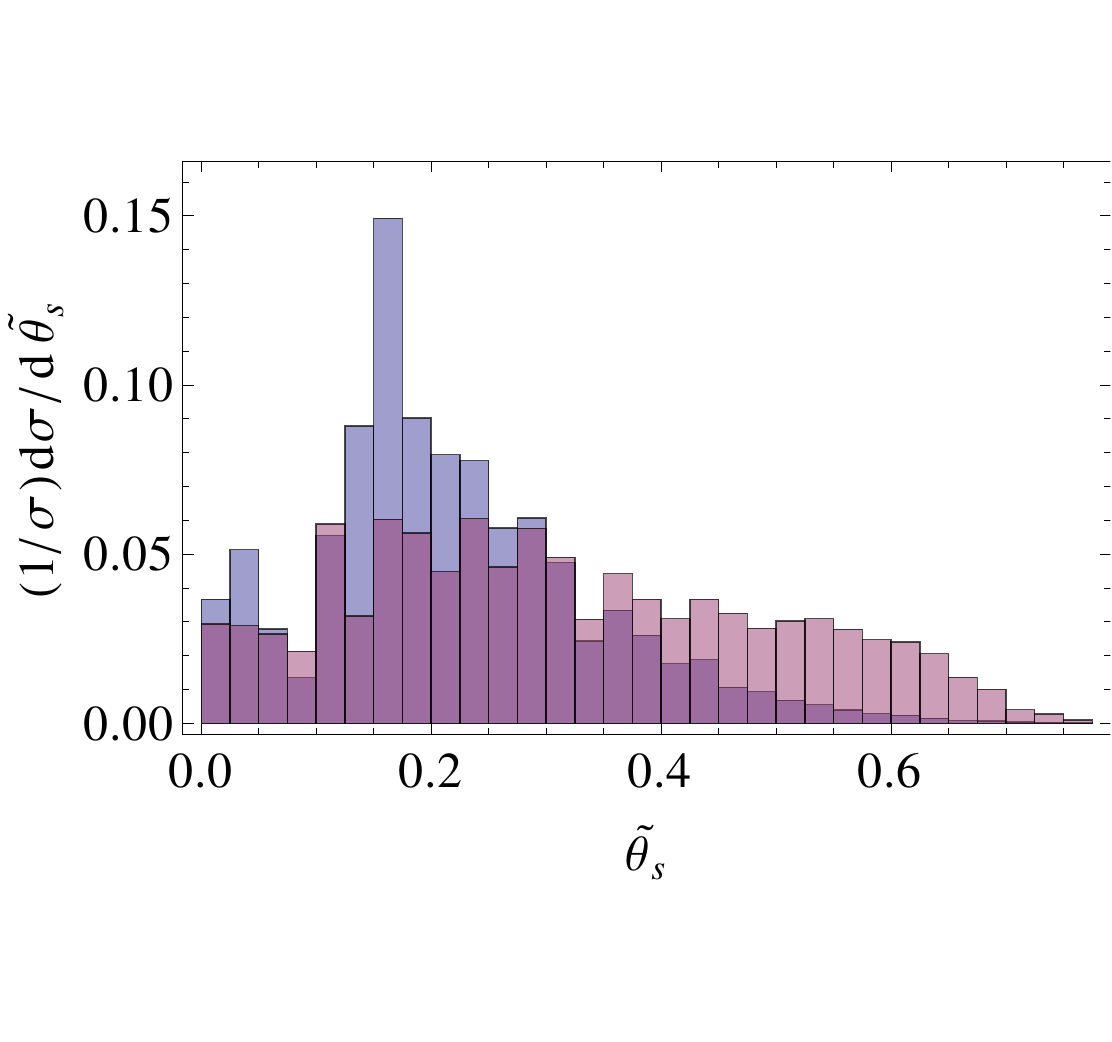}\\

\end{tabular}
\end{center}
\caption{A selection of various template variable distributions for Higgs jets (blue) and QCD background (purple) at the LHC. Events satisfy selection cuts and the Higgs mass window cut, $110 \GeV < m_J<130\GeV$. Horizontal axes are in radians or dimensionless units as appropriate, and vertical axes
are in arbitrary units with signal and background normalized to the same area.}\label{angles}
\end{figure}  
 
 In the Higgs decay to a quark-antiquark pair and a gluon,
$h\rightarrow q\bar{q}g$, we expect events where the gluon is soft
to be predominant. 
In the boosted frame,
this radiation appears dominantly within an angular region spanned by
the dipole formed by the quark and the antiquark \cite{colliderbook}. 
In contrast, in the perturbative expansion, jets initiated by quark or gluon radiation would have a color connection with the rest of the event resulting in a bias for large angle soft gluon emission towards other jets in the event or the beam.
 One can take advantage of this in order to focus on regions in
the phase space where Higgs decay events are more likely to dominate
over QCD events ({\it e.g.} with $ggg$ final state).

The variables $r_\theta$ and $\bar \theta$ are designed to measure the difference in angular ordering in our peak templates between the Higgs boson signal and the QCD background. 
We note that both $\bar{\theta}$ and $\theta_{12}$ for three-body
templates are somewhat analogous to $\theta_s$ for two-body
templates.


\section{Higgs tagging performance}\label{Sec:Higgs_tagging_performance}

In this section, we investigate the tagging efficiencies for Higgs jets and the mistag rates for
QCD jets using template overlap.  We apply it to events generated by {\sc Pythia 8.150}\cite{pythia8}, {\sc
  Sherpa 1.3.0}\cite{sherpa} (with CKKW matching \cite{ckkw}), and {\sc Madgraph}\cite{madgraph} interfaced to {\sc Pythia 6} \cite{pythia6} (with
MLM matching\cite{mlm}) also. The event selection was described in Sec. \ref{event_discretization}. 

As seen in Sec. \ref{overlap_plots}, both two- and thee-body template
overlap have substantial discriminating power. This can also be seen
in the scatter plots, shown in Fig. \ref{scatterOv},  of $Ov_2$
and $Ov_3$ for Higgs signal (left panel) and dijet production (right
panel). While the signal events cluster around the upper right corner
of the plot, most QCD jet events are localized diagonally opposite in
the lower left. It follows immediately that making tight cuts on each observable, by drawing a rectangular window in the upper right corner of the scatter plot, makes a good discriminator to separate signal from background.

\begin{figure}[hptb]
\begin{center}

\begin{tabular}{cc}

\includegraphics[width=.9\hsize]{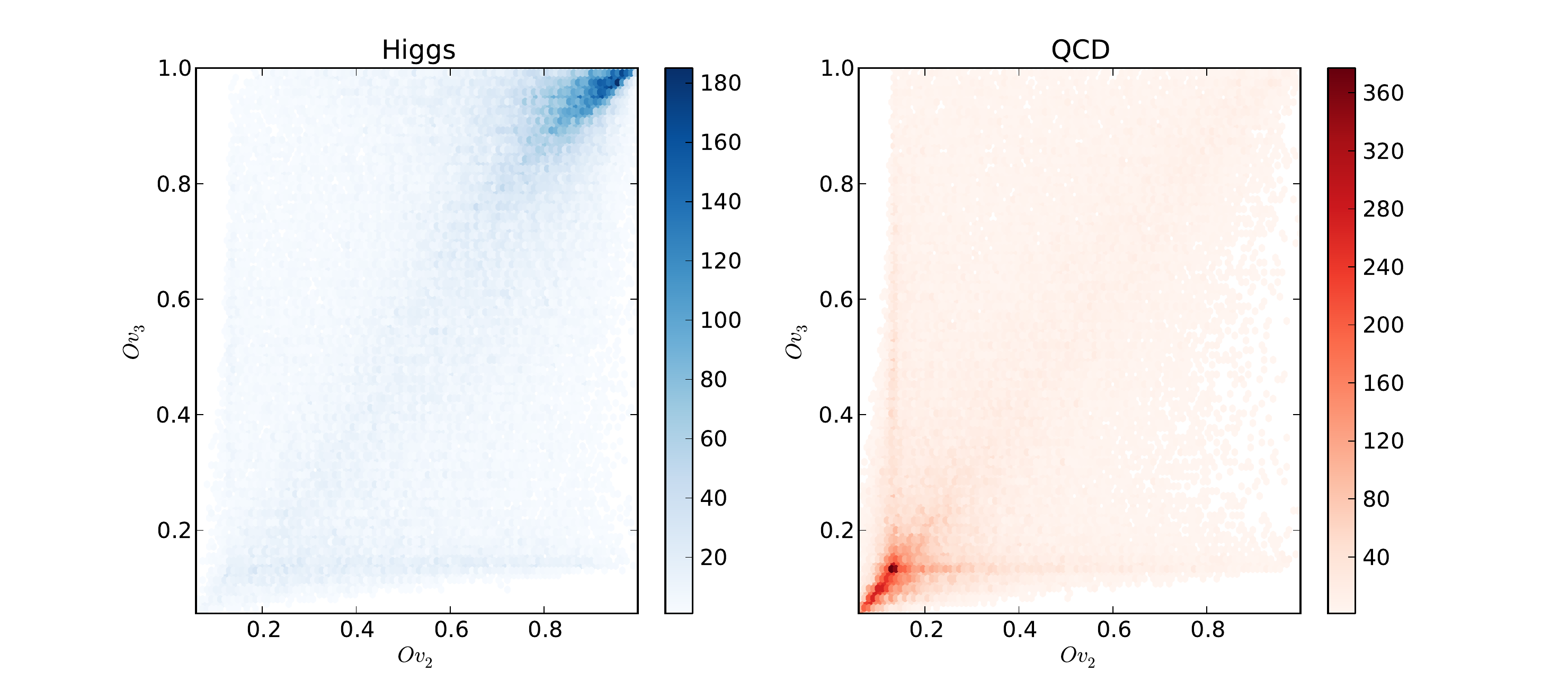}

\end{tabular}

\caption{
Density plots of 2-body overlap  {\it vs.}\ 3-body overlap for boosted
Higgs and QCD jets with $R=0.7$ and same number of events (20000).  
}\label{scatterOv}
\end{center}

\end{figure} 

We now assess the additional discriminating power offered by the template variables discussed in section \ref{Sec:Template_variables_methods}. 
In the simple analysis of this paper, we only look at the effect that a simple cut or window has on the number of signal and background events, leaving the use of more sophisticated methods for future work. We did not find that any of the template variables were qualitatively better than the planar flow cut. On the other hand, we found that planar flow and the $\bar \theta$ variable were somewhat complimentary. The two dimensional distribution $Pf$ vs $\bar \theta$ for both Higgs jets and QCD jets is shown in Fig. \ref{b_pf_comparison}. One can see that $Pf$ and $\bar \theta$ are approximately independent and, thus, drawing a contour to separate signal from background will do better than any single straight line. 
\begin{figure}[hptb]
\begin{center}
\begin{tabular}{cc}

\includegraphics[width=.9\hsize]{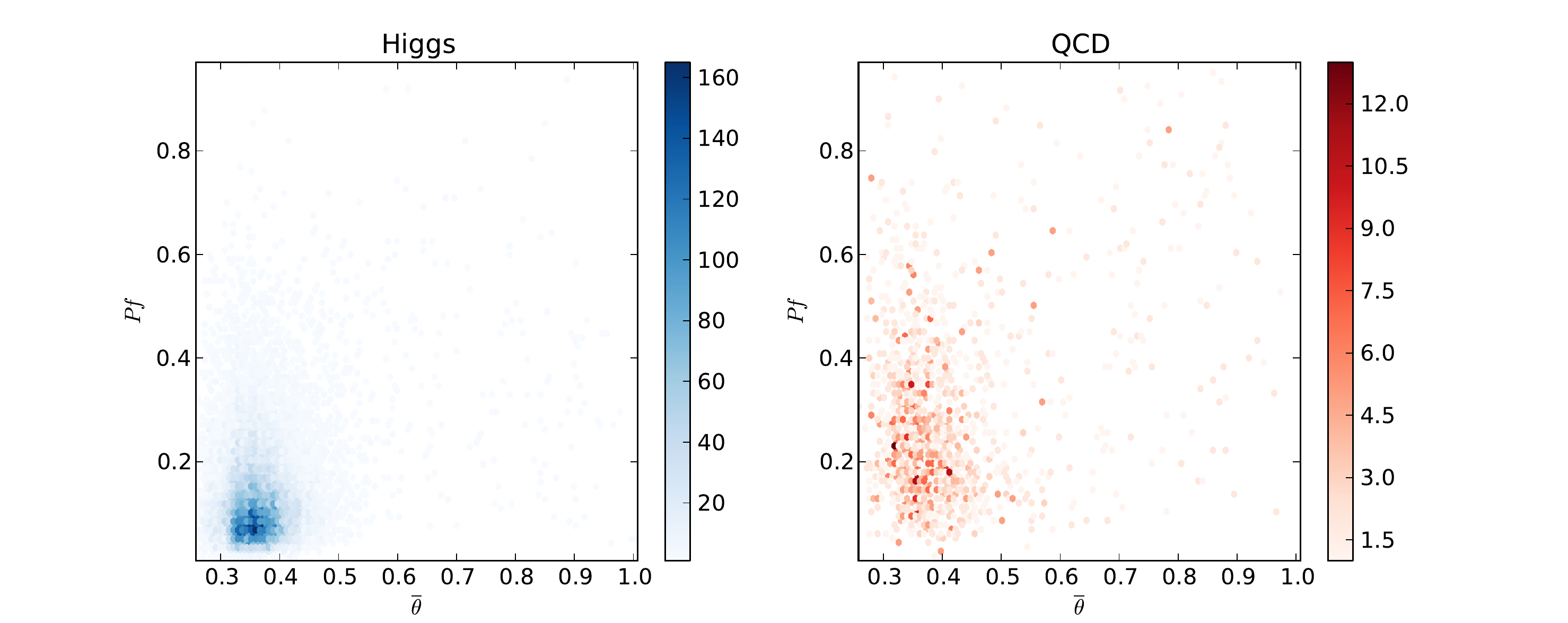}

\end{tabular}
\caption{ Density plots of $Pf$ vs $\bar \theta$ for anti-$k_T$ $R=0.7$ jets for Higgs signal (left) and QCD background (right) at the LHC with selection cuts and quality cut $Ov_2>0.8$ and $Ov_3>0.8$.}\label{b_pf_comparison}
\end{center}
\end{figure}

In Fig.~\ref{fake}, we summarize the boosted Higgs tagging efficiency
versus background rejection using the two overlaps observables $Ov_2$
and $Ov_3$ and the two variables $\bar \theta$ and
$Pf$. For a given lower cut on $Ov_3$ (denoted by the same color line
in each frame) the efficiency is controlled by the lower cut on two-parton overlap $Ov_2$. 
 Each point on one of these curves corresponds to a specific choice of $Ov_2$ at fixed $Ov_3$, and hence to
 the set of points within a rectangle that includes the upper right corners of the corresponding 
 scatter plots in Fig. \ref{scatterOv}.
 The results depend on the choice of $Ov_3$ cut, but it is clear that any cut above 0.8 leads to a 
 substantial increase in efficiency. 
  Without using properties of the data itself, such as jet shapes, it appears that purely template overlap variables can be used to remove significant amount of background. Planar flow cuts remove some of the background contributions, but $\bar \theta$ distributions also show some remaining discriminating power.
Once we combine the fake rate and efficiency from a jet mass cut (fake rate: $\sim 10\%$, efficiency: $\sim 70 \%$) with template overlap, $\bar \theta$ and planar flow, we find, for example, at efficiency of $10\%$, a fake rate of $0.05\%$ (with $Ov_3 >0.8$, $Pf< 0.2$ and $\bar \theta<0.4$).
 
 \begin{figure}[hptb]
\begin{center}

\begin{tabular}{cc}
\includegraphics[width=.33\hsize]{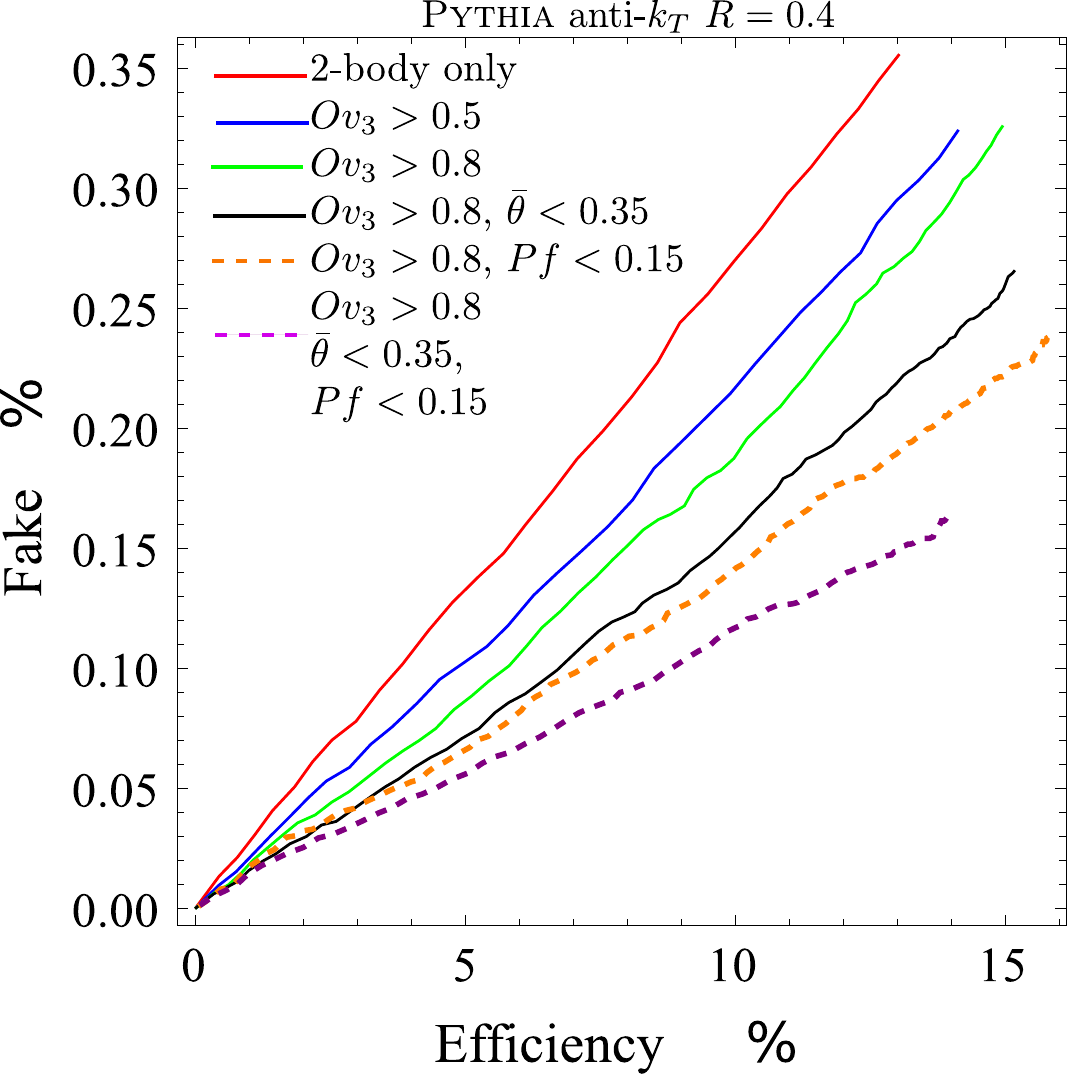}
\includegraphics[width=.33\hsize]{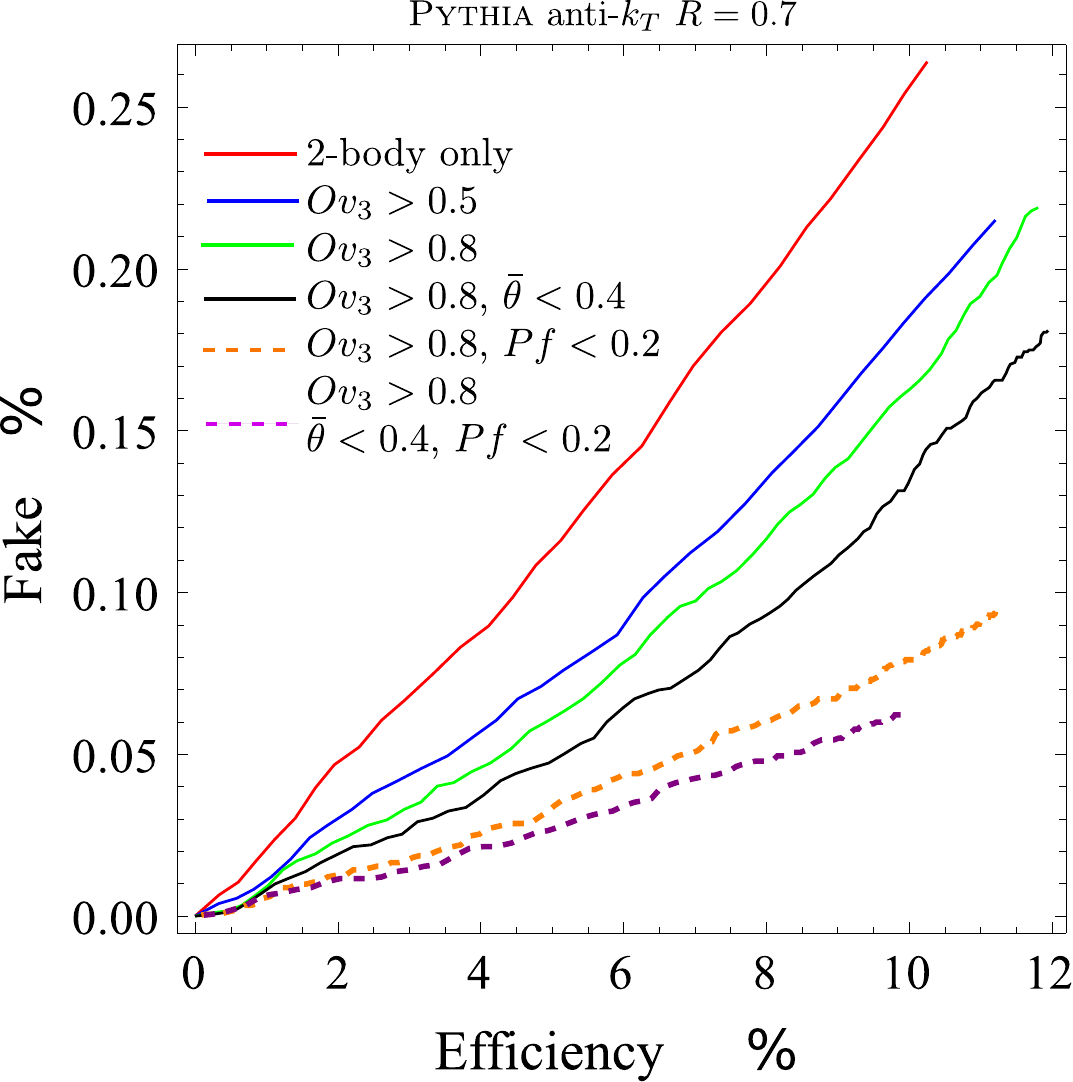}\\
\includegraphics[width=.33\hsize]{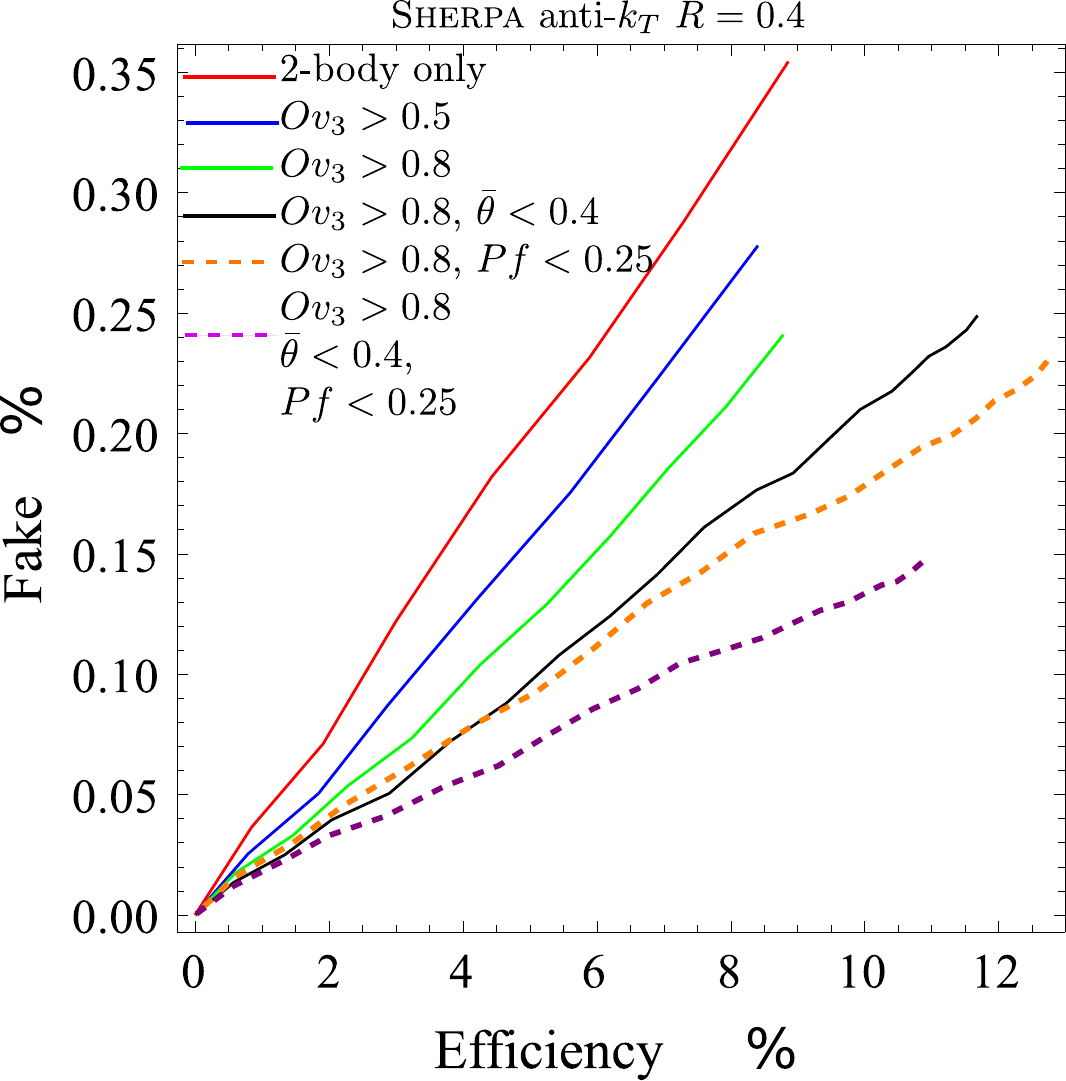}
\includegraphics[width=.33\hsize]{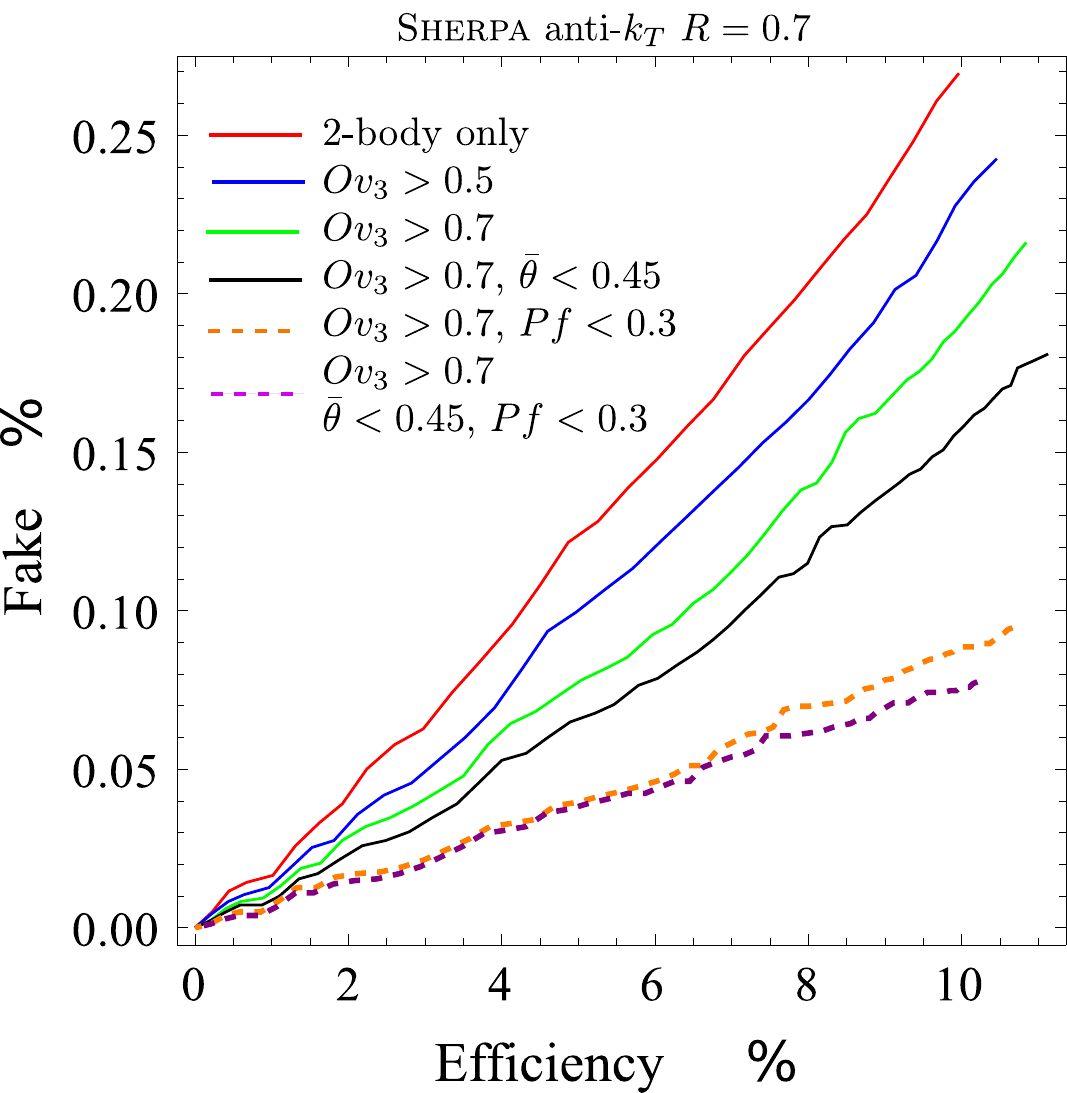}\\
\includegraphics[width=.33\hsize]{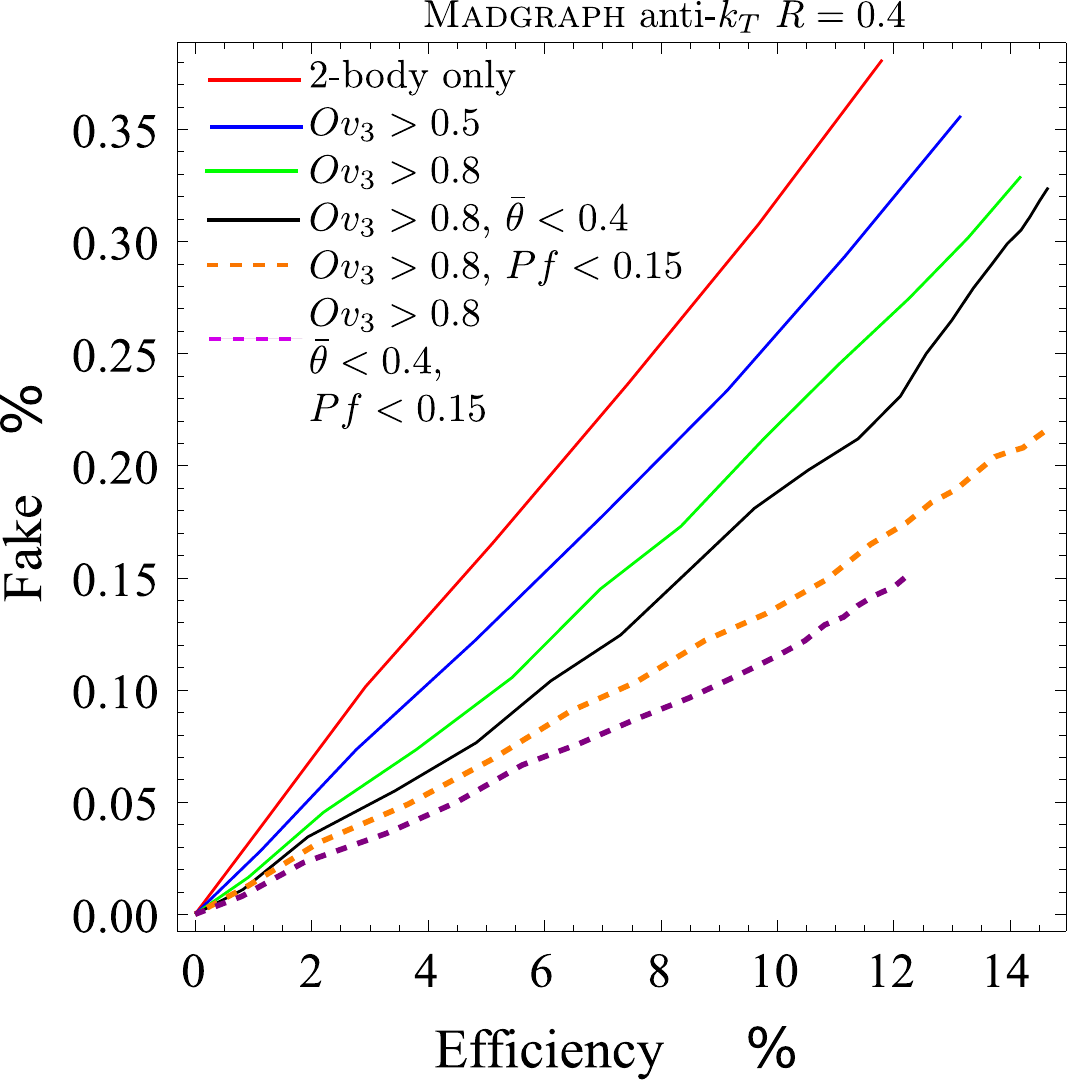}
\includegraphics[width=.33\hsize]{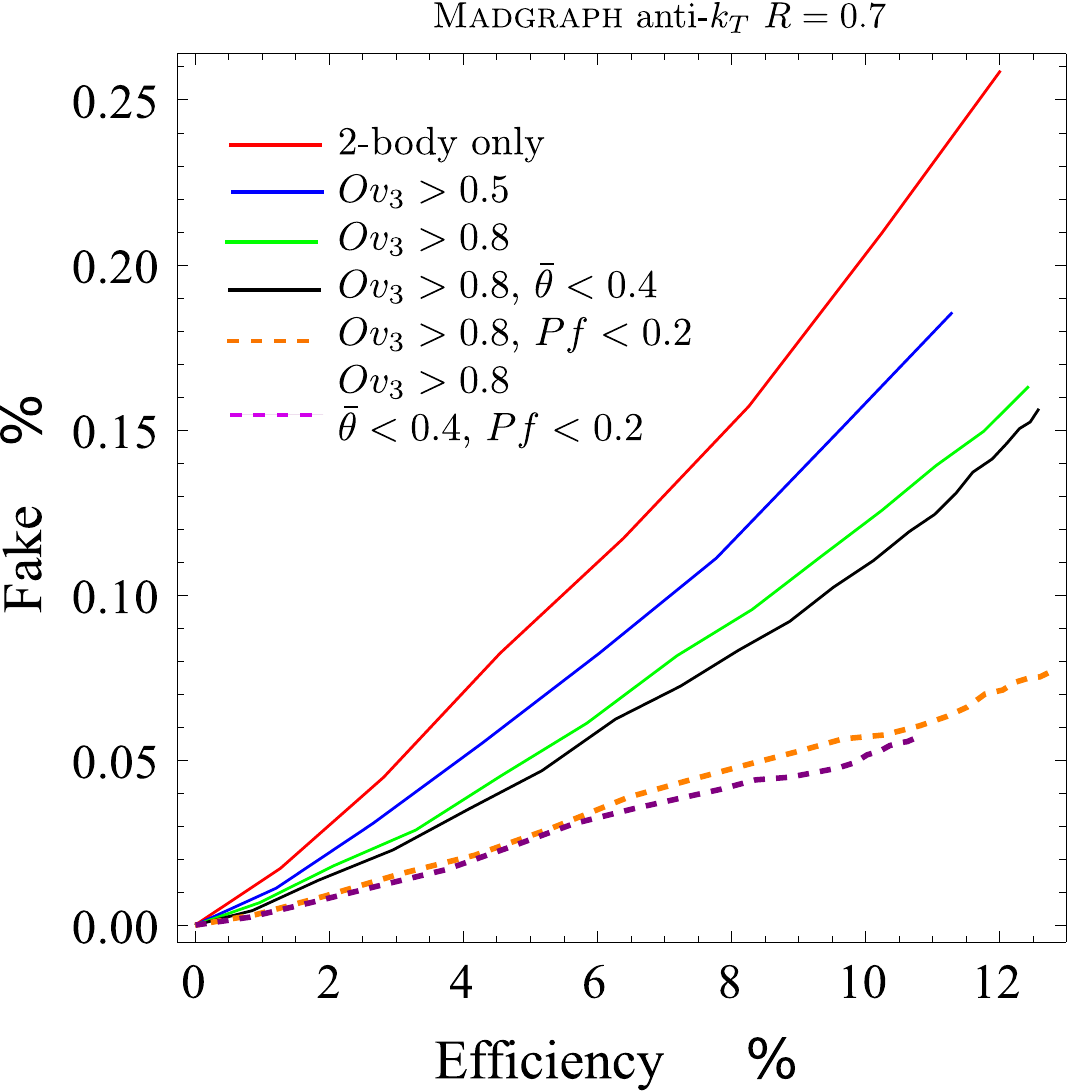}
\end{tabular}
\end{center}

\caption{Comparison of fake rate \textit{vs.} signal efficiency with
  various cuts on three-body template overlap $Ov_3$ with Higgs jets
  and QCD jets. The curves represent variations on the minimal value
  of two-body template overlap $Ov_2$\@. The
  frames show results from different MC [top {\sc Pythia}, middle {\sc
    Sherpa}, and bottom {\sc MG/ME} with $R=0.4$ (left) and $R=0.7$ (right)], for a Higgs mass window selection criteria $110\GeV < m_J<130\GeV$ and $950\GeV < P_0<1050\GeV$. Both efficiency and fake rates decrease as we increase the overlap cut. The dashed curves denote the case when $Pf$ cut is implemented, while the solid curves have no $Pf$ cut. }\label{fake}
\end{figure} 

Our final results for the Higgs jet case are summarized in Tables~\ref{tab:jetMass04} and \ref{tab:jetMass07} for the three event generators and $R=0.4,0.7$, respectively, that result from including these simple, naive one-dimensional cuts in $Ov_2$, $Ov_3$, $\bar \theta$ and $Pf$ at fixed signal efficiency of $S= 10\%$. 
 It is evident from the numbers presented that the template overlap method works well for events generated by any of the MC generators.
In each case, we find a large
enhancement of signal compared to background, typically of the order of fifteen or more. 
Taking into account the rejection of QCD jets by imposing a mass window, these numbers (for a single massive jet) are multiplied by
factors of ten to twenty. The efficiency for finding a jet within the
Higgs mass window is small for $R=0.4$, since a small cone is unlikely
to capture all of the QCD  
radiation in Higgs decay. For the bigger cone size $R=0.7$, the efficiency is increased roughly by a factor of 2.

For comparison, we have applied the BDRS Higgs tagger~\cite{Butterworth:2008iy} as implemented by {\sc FastJet 3}~\cite{fastjet}  to the {\sc Pythia} samples with the same jet radius and invariant mass window for QCD jets and Higgs jets. The BDRS Higgs tagger has two distinct parts: First, a mass-drop criteria to determine whether a jet contains substructure consistent with a Higgs boson decaying into bottom-antibottom quark pairs. If it passes the criteria, a jet is then filtered by reclustering it with a smaller jet radius $R_{filt}$ and only keeping the 3 hardest of these subjets as the pieces of the final filtered jet. One can then place cuts on the mass of the filtered jets, $m_{filt}$, to further reject background from QCD jets.  Our results can be seen in  Fig.~\ref{comparison2BDRS}, which compares the Template Overlap method to the BDRS Higgs tagger in a simple, non-optimized test. We see that the template-based approach and the BDRS Higgs tagger perform very similarly. While not an optimal comparison, this should serve as a useful point of reference to establish the effectiveness of Template Overlap.

 \begin{figure}[hptb]
\begin{center}

\begin{tabular}{cc}
\includegraphics[width=.5\hsize]{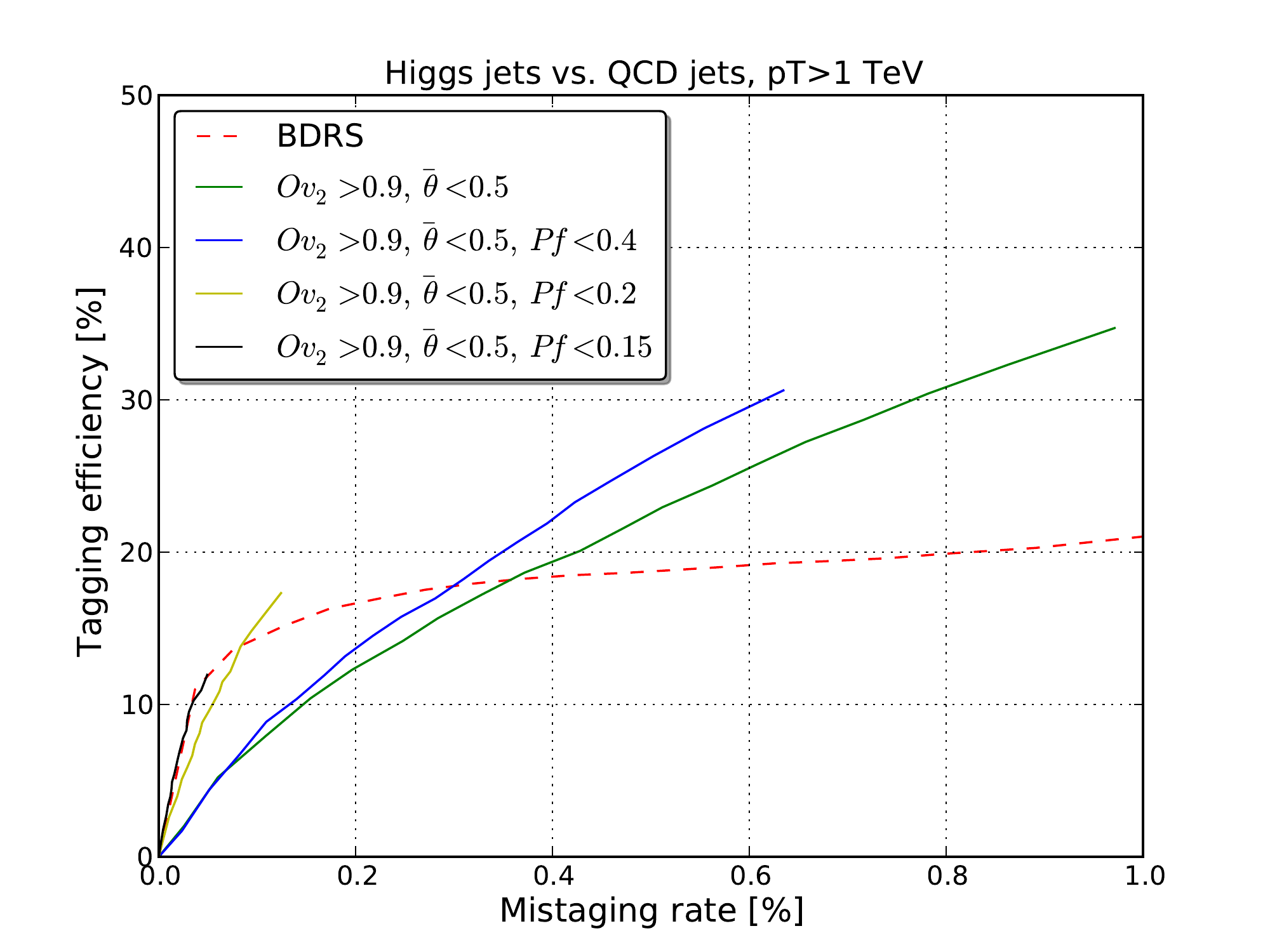}
\end{tabular}
\end{center}

\caption{Efficiency versus mistag rate for the  Template Overlap method and the  BDRS Higgs tagger.  The filled curves represent variations on the minimal value
  of three-body template overlap $Ov_3$\@, while the dashed curve corresponds to a varying mass window for the filtered jet mass, $m_{filt}$. The
results are from {\sc Pythia 8} with $R=0.7$ and same selection criteria as in the template case above.  }\label{comparison2BDRS}
\end{figure} 
 
 \begin{table}[htb]
\begin{center}
\begin{tabular}{c |c c|c c} \hline\hline
\rule{0pt}{1.2em}
MC &  \multicolumn{2}{c|}{Jet mass cut only} &
\multicolumn{2}{c}{ Mass cut + $Ov$ +$\bar \theta$+ $Pf$}  \cr
&   Higgs-jet efficiency~[\%] & fake rate~[\%] &  Higgs-jet efficiency~[\%] & fake rate~[\%] \cr
\hline {\sc Pythia 8}  &~$60$& $5$ &$10$& $0.10$
\\
MG/ME &~$60$& $5$& $10$& 0.10
\\
{\sc Sherpa} &~$50$& $5$& $10$& $0.10$
\\
\hline\hline
\end{tabular}
\caption{\label{tab:jetMass04} 
Efficiencies and fake rates for jets  with $R=0.4$ (using anti-$k_T$: $D=0.4$), 950 GeV$\le P_{0} \le$1050 GeV, 110 GeV$\le m_J \le$130 GeV and  $m_{H}=120$ GeV.  
 The left pair of columns shows efficiencies and fake rates found by imposing the jet mass window only.   The right
 pair takes into account the effects of cuts in both $Ov's$, $\bar \theta$ and $Pf$ in addition to the mass window.
 For the different MC simulations,
we have imposed various cuts on $Ov$, $\bar \theta$ and $Pf$ variables: for {\sc Pythia v8}~\cite{pythia8}  $Ov_2\ge 0.8$, $Ov_3\ge 0.8$, $\bar \theta<0.4$  and $Pf < 0.2$, for {\sc MG/ME}~\cite{madgraph} interfaced to {\sc Pythia v6}~\cite{pythia6}(with MLM matching~\cite{mlm}) $Ov_2\ge 0.8$, $Ov_3 >0.8$, $\bar \theta<0.4$ and $Pf< 0.2$
 and for {\sc Sherpa}~\cite{sherpa} (with CKKW matching~\cite{ckkw}) $Ov_2\ge 0.7$, $Ov_3 >0.7$, $\bar \theta<0.45$ and $Pf< 0.3$.
}

\end{center}
\end{table}
\begin{table}[hptb]
\begin{center}
\begin{tabular}{c |c c|c c} \hline\hline
\rule{0pt}{1.2em}
MC &  \multicolumn{2}{c|}{Jet mass cut only} &
\multicolumn{2}{c}{ Mass cut + $Ov$ +$\bar \theta $+ $Pf$}  \cr
&   Higgs-jet efficiency~[\%] & fake rate~[\%] &  Higgs-jet efficiency~[\%] & fake rate~[\%] \cr
\hline {\sc Pythia 8}  &~$70$& $10$ &$10$& $0.05$
\\
MG/ME &~$70$& $10$& $10$& 0.05
\\
{\sc Sherpa} &~$60$& $10$& $10$& $0.05$
\\
\hline\hline
\end{tabular}
\caption{\label{tab:jetMass07} 
Efficiencies and fake rates for jets  with $R=0.7$ (using anti-$k_T$: $D=0.7$), 950 GeV$\le P_{0} \le$1050 GeV, 110 GeV$\le m_J \le$130 GeV and  $m_{H}=120$ GeV.  
 The left pair of columns shows efficiencies and fake rates found by imposing the jet mass window only.   The right
 pair takes into account the effects of cuts in both $Ov's$, $\bar \theta$ and $Pf$ in addition to the mass window.
 For the different MC simulations,
we have imposed various cuts on $Ov$, $\bar \theta$ and $Pf$ variables: for {\sc Pythia v8}~\cite{pythia8}  $Ov_2\ge 0.8$, $Ov_3\ge 0.8$, $\bar \theta<0.4$  and $Pf < 0.2$, for {\sc MG/ME}~\cite{madgraph} interfaced to {\sc Pythia v6}~\cite{pythia6}(with MLM matching~\cite{mlm}) $Ov_2\ge 0.8$, $Ov_3 >0.8$, $\bar \theta<0.4$ and $Pf< 0.2$
 and for {\sc Sherpa}~\cite{sherpa} (with CKKW matching~\cite{ckkw}) $Ov_2\ge 0.7$, $Ov_3 >0.7$, $\bar \theta<0.45$ and $Pf< 0.3$.
}
\end{center}
\end{table}

 We present these results for demonstration purposes only,
 and have not carried out a systematic study of how to maximize rejection power.
 Because the template method naturally provides scatter plots like
 those in Figs.\ \ref{scatterOv} and  \ref{b_pf_comparison}, we can  imagine optimizing cuts on the data.
 We may also investigate improvements in the overlap functional Eq.\ (\ref{Nparticletemplate}).

\section{Conclusions}\label{Sec:Conclusions}

The template overlap method has the ability to maximize the physics reach for massive jets by matching jet energy flow with that of a boosted partonic decay. Template overlap is a particularly interesting observable, since it directly measures the N-prong nature of a jet. In this paper, we have proposed and tested several modifications to the template overlap method as implemented in \cite{Almeida:2010pa},
combining two- and three-body template states to the analysis of boosted Higgs.
Using several MC simulated samples, we have shown that template
overlap offers the promise of a successful boosted Higgs tagger,
validating the preliminary study in \cite{Almeida:2010pa}. We have
demonstrated how the inclusion of three particle templates allows us
to test the influence of gluon emission and color flow, through their
effect on energy flow, and have illustrated its use through the
construction of several partonic template observables. 
Different event
generators give moderately different averages for our template
overlaps. We nevertheless find in each case excellent and similar
rejection power. For the Higgs case studied in this work we get a rejection power of order 1:200 when combined with a jet mass cut, with sizable efficiencies.

We should emphasize that our event selection was chosen in a
kinematical regime that at present is unrealistic for the
LHC. However, our findings should serve as a proof of concept for many
of the ideas, and, based on ongoing research \cite{WIP,qcd_templates},
we expect an extended phenomenological analysis to deliver similar
qualitative behaviour in terms of rejection power.  The fact that the
rejection powers are strong is encouraging and should help motivate
possible modifications and improvements of the template method with
different analyses in mind. We have not yet made use of $b$-tagging, which is a
natural extension to the method. Since template overlap cuts are
independent of $b$-tagging, we expect the use of $b$-tagging  to result in significant
improvements in background rejection~\cite{WIP}. 
Additional discrimination power would be possible
using multivariate techniques, and we leave them for future study. Another point that we have not investigated in detail and that might be interesting in a future
study is how it is possible to fix some of the parameters in an optimal way given the properties of the events we want to reconstruct.
As we have seen in Section~\ref{Sec:Higgs_tagging_performance}, the choice of the jet cone radius $R$
has a strong impact on the performance of the method;
and it might be interesting to see how $R$ could be adapted
dynamically when varying the jet selection
criteria~\cite{arXiv:0903.0392}. It is also worth noting that the template method is quite general and could be employed in a broader set of applications to both Standard Model and beyond the Standard Model physics.  As the LHC continues to explore the energy frontier of particle physics, template overlap provides us with an interesting tool for further development of jet substructure techniques.

\section*{Acknowledgments}

We benefited much from continuous discussions of this and related subjects with R.~Alon, S. Frixione,  A.~Mitov, G.~Salam and D.~Soper.  J.J. is most grateful for the hospitality of the CERN Theory Group, and O.E. would like to 
thank the Weizmann Institute of Science for their hospitality, while part of this work was carried out. The work of O.\ E.\ and G.\ S.\ was supported by the
National Science Foundation, grants PHY-0653342 and PHY-0969739. GP is supported by the GIF, Gruber foundation, IRG, ISF and Minerva.

\section*{Appendices}

\appendix

\section{Kinematic of three-body decays}

We consider the process $p p \to H X \to  q\,\bar q\,g X$ of
Fig.~\ref{higgs_kinematics}, where $q$ and $\bar q$ are quark and antiquark and $g$ is a gluon. 
 In the approximation where the Higgs boson is exactly on-shell, the cross section for this process can be written in a factorized form as
\be
d\sigma_{pp\to X q\bar q g } =d\sigma_{pp\to X H} \frac{d {\rm \Gamma}_{H\to q\bar q g}}{\Gamma_0} \label{fullxsection}.
\ee
\begin{figure}[ht]
  \begin{center} 
\includegraphics[width=7cm]{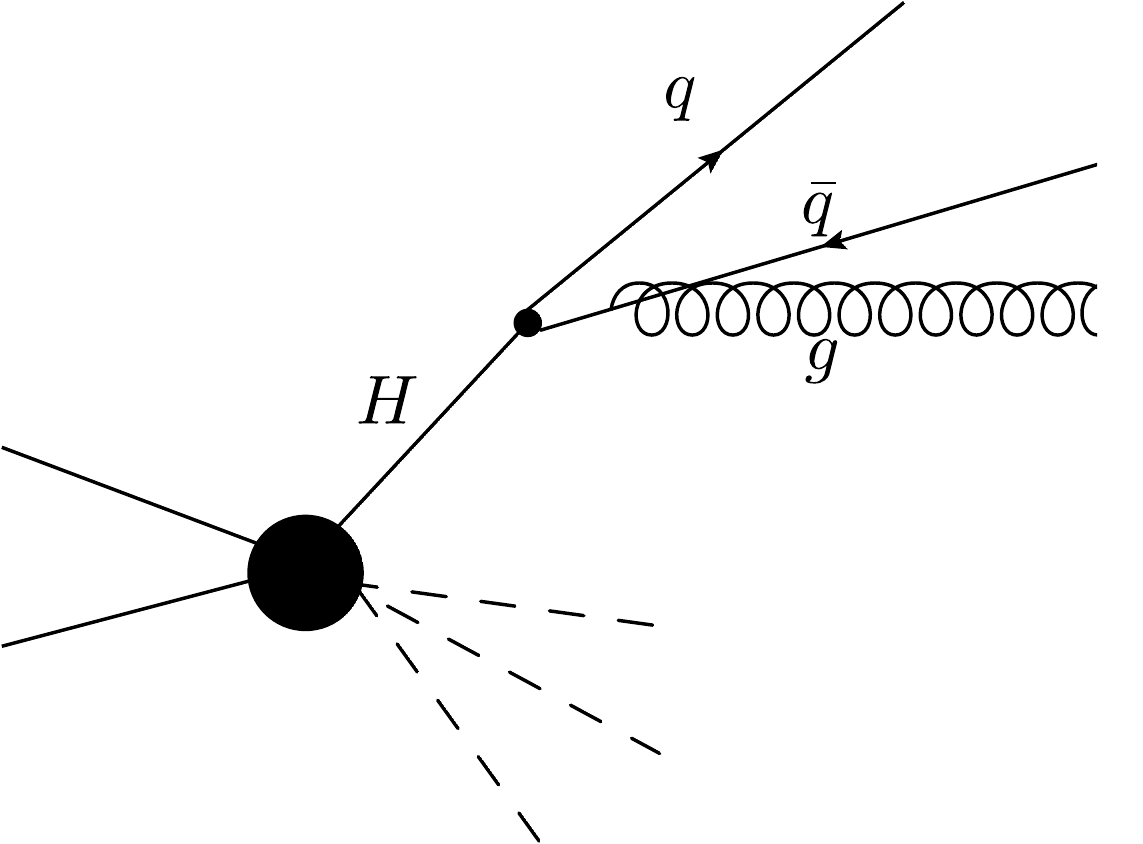} 
\end{center}
\caption{A schematic diagram of a Higgs production process and decay. For our study, the Higgs boson is assumed to be exactly on-shell.}\label{higgs_kinematics}
\end{figure}
It follows that the rest frame of the Higgs is identical to the center-of-mass frame of its decay products. In that frame the three final state particles will lie in a plane. The phase space can therefore be specified by giving three Euler angles $(\psi, \theta,\phi)$ that specify the  orientation of the  final system relative to the Higgs boson and fractional energy variables
\begin{equation}
{{x_i =  {E^*_i \over  m_H/2}}} = {2 p_i\cdot q\over m^2_H},\label{energy_fractions}
\end{equation}
with $E^*_i$ the CM energy, and 
\begin{equation}
{{ 0<x_i <1}} \ .
\end{equation}
Here, $q^\mu$ is the Higgs momentum with $q_{\mu} q^{\mu} = m^2_H$, $p_i^\mu$  are the momenta of the outgoing partons $(q,\bar q,g)$\@.
Energy conservation gives $\sum x_i =2$, which implies that only two of the $x_i$ are independent.

Let $\theta_{ij}$ be the angles between the momenta of the partons $i$ and $j$. One can relate the angles between the momenta to the energy fractions defined by
Eq.~(\ref{energy_fractions}). Momentum conservation gives the relation between the energy
fractions of the massless decay products and the angles between their
momenta in the rest frame of the Higgs,
\be (1-\cos\theta_{ij})\ =\ \frac{2(1-x_k)}{x_ix_j} \ , \quad i\neq
j\neq k \ . \ee
We can also obtain the angles between the boosted momenta in any frame,
\be (1-\cos\theta_{ij})\ =\ \frac{m^2_H}{2E_iE_j}(1-x_k) \ . \ee
The energies $E_i$ after a Lorentz boost ${ \vec{\gamma}}= \gamma \hat{n}$ are given by 
\begin{eqnarray}
E_1 &=& \frac{1}{2}(1+\beta n_1)\gamma m_Hx_1 \ \label{boostedE1}, \\ 
E_2 &=& \frac{1}{2}(1+\beta n_1\cos\theta_{12}-\beta
n_2\frac{2\sqrt{S}}{x_1x_2})\gamma m_Hx_2 \ , \label{boostedE2} \\
E_3 &=&
\frac{1}{2}(1+\beta n_1\cos\theta_{13}+\beta
n_2\frac{2\sqrt{S}}{x_1x_3})\gamma m_Hx_3 \ \label{boostedE3}
\end{eqnarray}
and we have defined,
\be
S \equiv  (1-x_1)(1-x_2)(1-x_3) \label{sphericity}.
\ee
Here, the boost direction is given by a unit vector in terms of two
Euler angles,
\be
\hat{n}=(n_1,n_2,n_3)=(\cos\phi\sin\theta,\sin\phi\sin\theta,\cos\theta).
\ \ee
In our conventions, the initial $z$-axis is perpendicular to the
plane of the decay in the rest frame of the decaying particle, and we
chose the $x$-axis to point in the direction of the quark.

\section{Templates at NLO}

As discussed in Sec.\ \ref{meopt}, templates were generated in three-particle phase space  with a density that reflects the NLO differential decay rate for Higgs decay, Eq.\ (\ref{3_body_ps}).   To be counted as a three-particle template, each pair of partons in the final state should have a fractional invariant mass, $y_{ij}>y$, in Eq.\ (\ref{invariant_mass_cut}), where in this study we have taken $y=0.05$.    As shown in Fig.\ \ref{validate}, the output of partonic level Monte Carlos suggests that about 20\% of events generated in this way will pass this cut.   In fact, we can confirm this estimate and calculate NLO $y$ dependence analytically, by integrating the differential rate, Eq.\ (\ref{3_body_ps}) subject to $y_{ij}>y$ for all pairs $i,j$.    Following familiar notation \cite{colliderbook}, this result divided by the total NLO  decay rate is labelled $f_3$, the three-jet fraction.   The explicit expression is
\begin{multline}\label{jade3}
f_3=\frac {C_F \alpha_s} {4 \pi}\left [-4\text{Li}_ 2\left (2 + \frac {1} {y - 1} \right) + 
   6 y\log\left (\frac {1} {y^2} - \frac {3} {y} 
      2 \right)   \right. \\\left.  + \log (y)\left (-4\log\left (2 y^2 - 3 y + 
         1 \right) +   4\log (y) + 3 \right) +  (2 y - 1) (5 y - 
      7)   \right. \\\left. +  \log (1 - 2 y) (4\log (1 - y) - 3) - 
   6\tanh^{-1} (1 - 2 y) \right]\, ,
\end{multline}
We can regard the remaining fraction of events as a two-jet fraction, labelled
\be
f_2 = 1- f_3\, . \label{jade2}
\ee
The fractions $f_2$ and $f_3$ are shown in Fig. \ref{jade_br} as a function of $y$. The points correspond to the reference value, $y=0.05$ used in our study.
\begin{figure}[ht]
  \begin{center} 
\includegraphics[width=7cm]{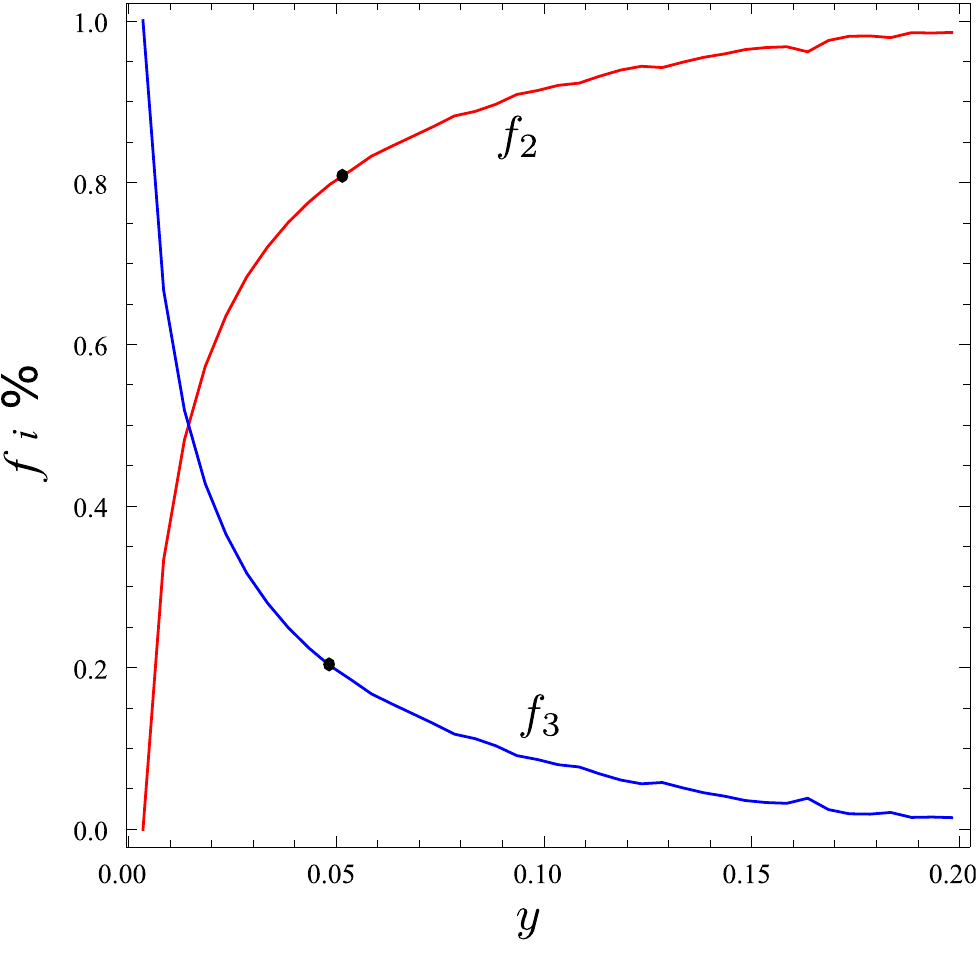} 
\end{center}
\caption{The values of $f_3$ and $f_2$ from
  Eq. (\ref{jade3})-(\ref{jade2}). Notice that the choice $y=0.05$
  corresponds to roughly $f_2=0.8$ and $f_3=0.2$\@.} \label{jade_br}
\end{figure}

\pagebreak

\end{document}